\documentclass[5p,final]{elsarticle}
\biboptions{round,sort,authoryear}
\usepackage{etex}
\usepackage{fixltx2e}
\usepackage{pictex}
\usepackage{color}
\usepackage{url}
\usepackage{amsmath}
\usepackage{rotating}
\usepackage{mdwlist}

\newenvironment{flushenum}{
\begin{enumerate}
  
  \setlength{\leftmargin}{0pt}
  \setlength{\leftmargini}{0pt}
  \setlength{\leftmarginii}{0pt}
  \setlength{\leftmarginiii}{0pt}
  \setlength{\leftmarginiv}{0pt}
  \setlength{\labelwidth}{1em}
  \setlength{\itemindent}{0em}
  \setlength{\listparindent}{0em}
}{\end{enumerate}}
\definecolor{orange}{rgb}{1,0.5,0}
\usepackage{xr}
\begin{document}
\begin{frontmatter}
\title{Bias in Estimators of Archaic Admixture}
\author[au1]{Alan R. Rogers\corref{cor1}}
\ead{rogers@anthro.utah.edu}
\author[au2]{Ryan J. Bohlender}
\ead{ryan.bohlender@gmail.com}

\cortext[cor1]{Corresponding author}
\address[au1,au2]{Dept.{} of Anthropology, 270~S 1400~E,
University of Utah, Salt Lake City, Utah 84112}

\begin{abstract}
This article evaluates bias in one class of methods used to estimate
archaic admixture in modern humans.  These methods study the pattern
of allele sharing among modern and archaic genomes. They are sensitive
to ``ghost'' admixture, which occurs when a population receives
archaic DNA from sources not acknowledged by the statistical
model. The effect of ghost admixture depends on two factors:
branch-length bias and population-size bias. Branch-length bias occurs
because a given amount of admixture has a larger effect if the two
populations have been separated for a long time.  Population-size bias
occurs because differences in population size distort branch lengths
in the gene genealogy. In the absence of ghost admixture, these
effects are small. They become important, however, in the presence of
ghost admixture.  Estimators differ in the pattern of
response. Increasing a given parameter may inflate one estimator but
deflate another. For this reason, comparisons among estimators are
informative. Using such comparisons, this article supports previous
findings that the archaic population was small and that Europeans
received little gene flow from archaic populations other than
Neanderthals. It also identifies an inconsistency in estimates of
archaic admixture into Melanesia.

\end{abstract}
\end{frontmatter}

\section{Introduction}
\label{sec.intro}

Forty years ago, William \citet{Howells:JHE-5-477} discussed the
origin of modern humans, emphasizing two extreme views. One of these,
which would now be called the multiregional hypothesis
\citep{Wolpoff:Multiregional}, held that modern humans evolved across
a broad front within a worldwide population held together by gene
flow. The other, which would now be called the replacement hypothesis
\citep{Stringer:S-239-1263}, involved ``a single origin, outward
migration of separate stirps, like the sons of Noah, and an empty
world to occupy, with no significant threat of adulteration by other
gene pools or even evaporating gene puddles''
\citep[p.~480]{Howells:JHE-5-477}. But Howells also considered a third
hypothesis, which also proposed expansion from a single point of
origin. This expansion, however, involved ``encounters between
populations of modern man and of other forms, with consequent gene
flow'' \citep[p.~492]{Howells:JHE-5-477}. This hypothesis has been
endorsed by various paleoanthropologists \citep{Brauer:OMH-84-327,
  Brauer:EMH-89-123, Smith:YPA-32-35, Trinkaus:ARA-34-207}. During the
past decade, it has also received support from genetics.

In the preceeding decade, geneticists were less supportive. At that
time, human evolutionary genetics dealt mainly with mitochondrial DNA
(mtDNA), which is remarkably homogeneous in modern human samples.
\citet{Stoneking:EA-2-60} argued that the mtDNA of Neanderthals ought
to lie well outside the narrow range of variation seen in modern human
samples. The absence of such divergent mtDNAs argued that their
frequency within the human species must be low. Yet as Stoneking
observed, this did not refute the hypothesis of archaic
admixture. Introgressed archaic mtDNAs might simply have been lost by
genetic drift.  \citet{Nordborg:AJH-63-1237} developed a model of this
process, which showed that mitochondrial data have low power to detect
archaic admixture.

Since the late 1990s, the field has relied increasingly on nuclear
DNA. Because unlinked loci provide essentially independent replicates
of the evolutionary process, the nuclear genome provides far greater
power to detect admixture. A variety of statistical methods has been
developed. Some rely on information in the site frequency spectrum
\citep{Eswaran:JHE-49-1, Yang:MBE-29-2987}. Others are based on
linkage disequilibrium \citep{Wall:G-154-1271, Wall:MBE-26-1823,
  Wall:COG-16-606, Plagnol:PLO-2-e105, Moorjani:PLO-7-e1001373,
  Hammer:PNA-37-15123, AbiRached:S-334-89, Evans:S-309-1717,
  Mendez:MBE-29-1513, Cox:G-178-427}.

\begin{table}
  \caption{The site patterns studied in this analysis (with 0 and 1
    representing the ancestral and derived alleles); the
    gene tree implied by each pattern; and the counts $(I_{uv})$ of
    such sites for a San sample, $x$, a French sample, $y$, a
    Neanderthal sample, $n$, and a chimpanzee sample, $o$
    \citep[p.~S138]{Patterson:S-328-S129}.}
\label{tab.pat}
{\centering\begin{tabular}{cc@{}c@{}c@{}ccr@{$\;=\;$}l}
\hline
\multicolumn{5}{c}{Site}\\
\multicolumn{5}{c}{Pattern}\\ \cline{1-5}
 &$x$&$y$&$n$&$o$&Gene tree&\multicolumn{1}{c}{}&count\\
\hline\hline
$ny$& 0 & 1 & 1 & 0 &$((x,(y,n)),o)$ & $I_{ny}$&103,612\\
$nx$& 1 & 0 & 1 & 0 &$(((x,n),y),o)$ & $I_{nx}$&95,347\\
$xy$& 1 & 1 & 0 & 0 &$(((x,y),n),o)$ & $I_{xy}$&303,340\\
\hline
\end{tabular}\\}
\end{table}

Our focus here is on a different class of methods, which capitalizes
on the availablity of archaic DNA sequences. These methods infer
admixture from the frequency with which derived alleles are shared by
pairs of samples. In the most common pattern, the derived allele is
shared by genes drawn from closely related populations. Two samples
uniquely share a derived allele only if a mutation occurs in a
uniquely shared ancestor.  For example, at many of the loci in
table~\ref{tab.pat}, the derived allele is present only in the French
and African samples. These derived alleles arose in genes that were
ancestral to the French and African samples but not to the Neanderthal
or the Chimpanzee. Such sites are common in the data, because the
French and African populations are conspecific and thus share a
portion of their evolutionary history.

What then of the other two patterns, in which the derived allele is
shared by a Neanderthal and one of the two modern human samples? In
the absence of admixture, these site patterns can arise only through
incomplete lineage sorting. If random mating prevailed within the
population ancestral to humans and Neanderthals, these two patterns
ought to occur in equal frequencies \citep{Pamilo:MBE-5-568}. Yet in
table~\ref{tab.pat} the $ny$ pattern occurs more often than the $nx$
pattern. This excess supports the hypothesis of admixture between
Neanderthals and the ancestors of Europeans. Several published methods
use this principle to estimate the fraction of archaic genes in modern
populations \citep{Green:S-328-710, Reich:N-468-1053,
  Durand:MBE-28-2239, Reich:AJH-89-516, Meyer:S-338-222,
  Patterson:G-192-1065}.

These methods rely on the assumption of random mating in the ancestral
population. If instead that population were geographically structured,
with limited gene flow between geographic subdivisions, this could
result in biased frequencies such as those seen in table~\ref{tab.pat}
\citep{Slatkin:MBE-25-2241}. This hypothesis of ``ancestral
subdivision'' has been seen as an alternative to that of archaic
admixture \citep{Durand:MBE-28-2239, Eriksson:PNA-109-13956,
  Blum:MBE-28-889}. This issue is still contentious, with some authors
arguing that it has been refuted \citep{Yang:MBE-29-2987,
  Sankararaman:PLO-8-e1002947, Wall:G-194-199} and others that it has
not been properly tested \citep{Eriksson:MBE-31-1618}.

Whatever the outcome of this dispute, there are also other potential
biases. Several published estimators allow for gene flow from only one
archaic population. Estimates may be biased if the modern population
also received genes from other archaic populations, a phenomenon known
as ``ghost admixture'' \citetext{\citealp{Beerli:ME-13-827};
  \citealp{Slatkin:ME-14-67}; \citealp[p.~2240]{Durand:MBE-28-2239};
  \citealp{Harris:PLO-9-e1003521}}. Some estimators also assume that
population size has been constant throughout the human gene
tree. These estimators may be biased if populations have varied in
size.  In what follows, we explore the magnitudes of these biases.

\section{Methods}
\label{sec.methods}

Because this article is about bias, we focus on expected values and
ignore statistical uncertainties. Following
\citet[p.~2241]{Durand:MBE-28-2239}, we assume that admixture occurs
at discrete points in time. Between these events, the isolation of
populations is complete. Within populations, we assume mating is at
random.

\subsection{Population sizes and coalescent time scale}
\label{sec.popsize}

We use single upper-case letters, such as $X$ and $Y$, to label
individual populations. The notation $XY$ refers to the population
ancestral to $X$ and $Y$ but not ancestral to other sampled
populations.  The diploid sizes of $X$, $Y$, and $XY$ are written as
$N_X$, $N_Y$, and $N_{XY}$. The symbol $N_0$ represents the
diploid size of the the ancestral human population---the ancestors of
modern humans, Neanderthals, and Denisovans, but not of chimpanzees.

In this population, the hazard of a coalescent event between a single
pair of lineages is $1/2N_0$ per generation, and their mean
coalescence time is $2N_0$ generations
\citep{Hudson:OSE-7-1}. However, let us adopt a coalescent time scale,
with time units of $2N_0$ generations.  On this scale, the mean and
hazard are both unity, and the mutation rate is $U \equiv 2N_0u$,
where $u$ is the mutation rate per generation.

\begin{figure*}
{\centering\input{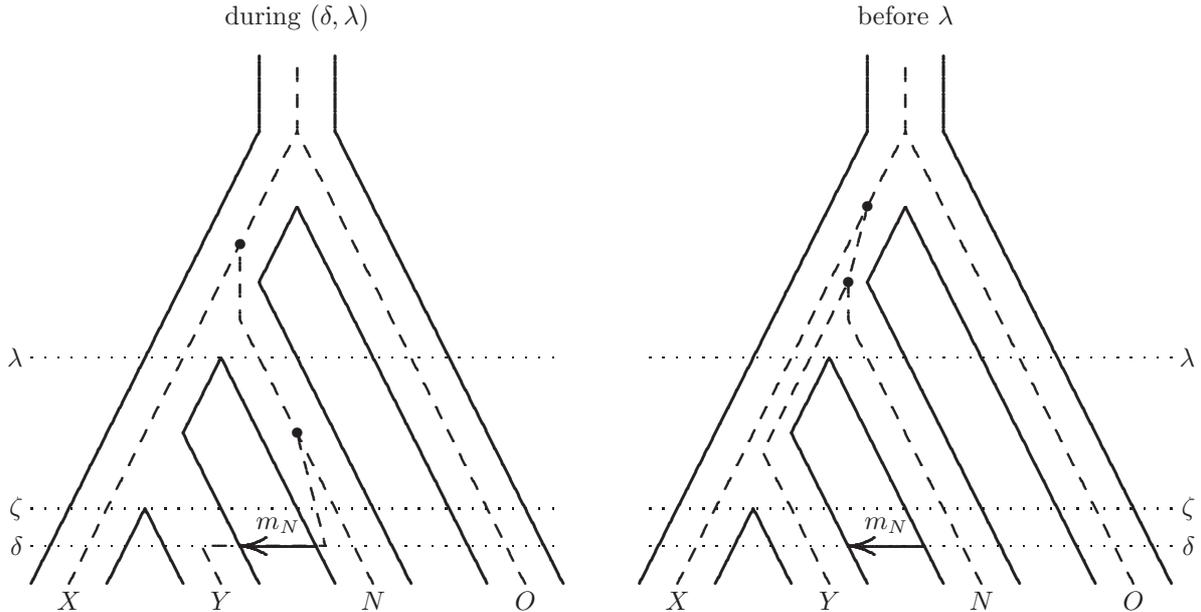}\\}
\caption{A population tree (solid lines) with two embedded gene trees
  (dashed lines). A fraction $m_N$ of the sites in population $Y$
  descend via gene flow from population $N$. Greek letters $\delta$,
  $\zeta$, and $\lambda$ refer to the time of the episode of gene
  flow, the separation of populations $X$ and $Y$, and the separation
  of $A$ from the two modern populations. Left panel: nucleotide
  carried by $y$ is derived by gene flow from $N$ and coalesces during
  $(\delta,\lambda)$ within population $N$. Right
  panel: nucleotide in $y$ is native and coalesces prior to $\lambda$
  within ancestral human population, $XYN$. Bullets delineate
  the branches along which mutation would generate site pattern $ny$.}
\label{fig.mytree}
\end{figure*}

We allow for changes in population size at branch points in the
population tree. Between branching points, we assume the population is
constant. For example, Fig.~\ref{fig.mytree} implies that population
$XY$ existed within the time interval $(\zeta, \lambda)$. Within this
interval, we assume that it had constant size $N_{XY}$. Let $K_{XY} =
N_{XY}/N_0$. In words, $K_{XY}$ is the size of population $XY$
relative to that of the ancestral human population.  For the duration
of population $XY$, the coalescent hazard for a single pair of
lineages is $1/K_{XY}$ per unit of coalescent time. Other ratios, such
as $K_X$ and $K_Y$, are defined similarly.

The ``survival function'', $S_{XY}^{(\zeta,\lambda)} \equiv
e^{-(\lambda-\zeta)/K_{XY}}$, is the probability that a pair of
lineages within $XY$ remain distinct throughout interval
$(\zeta,\lambda)$. The ``cumulative distribution function,''
$F_{XY}^{(\zeta,\lambda)} = 1-S_{XY}^{(\zeta,\lambda)}$, is the
probability that the pair coalesces within this interval.

For a pair of lineages within population $X$ at time $t$, $T_X^{(t)}$
is the expected coalescence time in units of $2N_0$ generations. It
depends not only on the size of population $X$, but also on the sizes
of populations ancestral to $X$. \ref{sec.T} explains how $T_X^{(t)}$
is calculated. If population size is constant throughout the
population tree, $K_X = T_X^{(t)} = 1$.

For numerical results, we assume the population sizes shown in
table~\ref{tab.psize}. The last line there is based on
\citet[table~2]{Gravel:PNA-108-11983}, who estimate that 932
generations ago, the European was of diploid size 1032. It then
expanded exponentially at a rate of 0.0038 per generation. Over this
interval, the harmonic mean population size would have been 3608. A
similar calculation, based on the estimates of Gravel et al.\ for East
Asia, gives a harmonic mean of 2446. Our own value of 3000 is a round
number midway between these results. This value of 3000 was then
scaled up or down to obtain lines 1--4 of table~\ref{tab.psize}, the
scaling ratios for lines 2--4 being chosen for consistency with
\citet[Fig.~4]{Prufer:N-505-43}, and that for line 1 for consistency
with \citet{Wall:G-163-395}.

\begin{table}
\caption{Assumptions about population sizes}
\label{tab.psize}
{\centering
\begin{tabular}{lrr}
           & Diploid\\
Population & size & $K$\\
\hline
Chimpanzee-human ancestor$^a$        & 15,000 & 6.50\\
Ancestor of moderns and archaics$^b$ &  2,308 & 1.00\\
Archaics$^b$                         &    577 & 0.25\\
Early modern humans$^b$              &  4,615 & 2.00\\
Europe or Asia$^c$                   &  3,000 & 1.30\\
\hline
\end{tabular}\\}
$^a$\citet{Wall:G-163-395}; 
$^b$\citet[Fig.~4]{Prufer:N-505-43};
$^c$\citet[table~2]{Gravel:PNA-108-11983}.
\end{table}

\subsection{Site patterns and their expected frequencies}
\label{sec.logic}

This section outlines the logic underlying the estimators that we
consider below. It was introduced by \citet{Green:S-328-710} and has
been used in many subsequent publications.

Consider a sample consisting of one haploid genome from each of four
populations: two modern human populations, $X$ and $Y$, one archaic
population, $N$, and an outgroup, $O$. For example, $X$ and $Y$ might
refer to the Yoruban and French populations, $N$ to ancient
Neanderthals, and $O$ to chimpanzee. We use lower case $(x, y, a, o)$
to refer to the genomes sampled from these populations. We restrict
attention to loci (nucleotide sites) at which two of the three human
samples carry the derived allele, 1, and the other sample carries the
ancestral allele, 0. By assumption, chimpanzee carries the ancestral
allele.

Suppose that we sample one haploid genome from each population.  The
gene genealogy of this sample will vary from locus to locus, but many
of these genealogies will have a topology similar to that of the
population. We assume this topology is $(((X,Y),N),O)$. Many gene
genealogies will however have different topologies, which may arise
through incomplete lineage sorting. If the most recent coalescent
event precedes the separation time, $\lambda$, of the archaic and
modern populations, then there are three possible genealogies:
$(((x,y),n),o)$, $(((x,n),y),o)$, and $((x,(y,n)),o)$. The right panel
of Fig.~\ref{fig.mytree} illustrates the last of these alternatives.
If the ancestral population mated at random, the three genealogies are
equally probable. In the absence of archaic admixture, two of these
genealogies---the ones inconsistent with the population tree---can
arise only by incomplete lineage sorting and should therefore be
equally frequent. But if there were gene flow from $N$ to $Y$, as seen
in the left panel of Fig.~\ref{fig.mytree}, we'd have an excess of
sites exhibiting genealogy $((x,(y,n)),o)$.

\begin{figure*}
{\centering\input{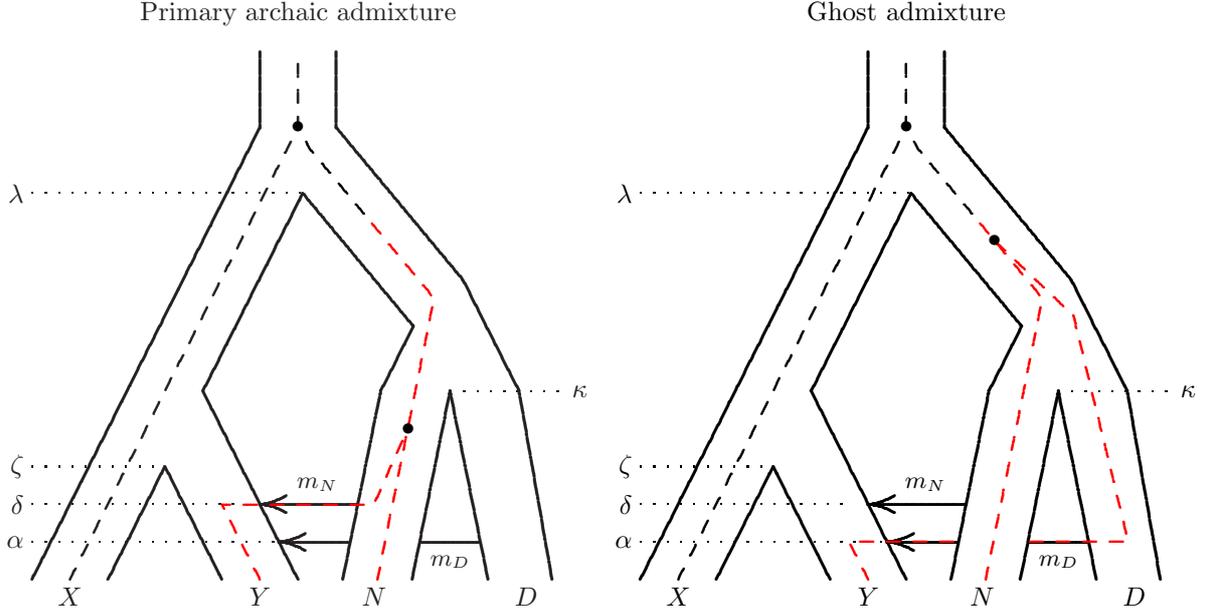}\\}
\caption{A population tree (solid lines) with embedded gene trees
  (dashed lines). In a sample from population $Y$, a fraction $m_D$ of
  nucleotide sites descend from archaic population $D$. Of the
  remaining fraction $1-m_D$, a fraction $m_N$ descend via gene flow
  from archaic population $N$. Greek letters indicate time variables,
  as explained in table~\ref{tab.time}. Bullets delineate branches
  along which mutation would generate site pattern $ny$. Red indicates
  portion of genealogy carrying derived allele. Left panel: nucleotide
  sampled from $Y$ comes from $N$ and coalesces during
  $(\delta,\lambda)$ within population \textit{ND}. Right panel:
  nucleotide sampled from $Y$ comes from $D$ and coalesces during
  $(\kappa, \lambda)$.}
\label{fig.xynd}
\end{figure*}

We can detect these gene genealogies only when a mutation occurs in
the common ancestor either of $(y,n)$, of $(x,n)$, or of $(x,y)$. For
example, the left panel of Fig.~\ref{fig.xynd} shows the case of a
site at which the nucleotide carried by $y$ arrived via gene flow
from population $N$. Tracing this lineage backward in time, it
coalesces with $n$ and then with $x$ at the points marked in the
figure by bullets. If a mutation occurred between these two coalescent
events, it would be shared by $y$ and $n$, but not by $x$ or
$o$. Thus, admixture from archaic population $N$ inflates the count of
sites that exhibit pattern $ny$. Our goal is to use this excess to
estimate the rate, $m_N$, of gene flow from $N$ into $Y$.

Such estimates may be biased for several reasons, one of which---ghost
admixture---is illustrated in Fig.~\ref{fig.xynd}. There, population
$Y$ receives gene flow not only from $N$, but also from a second
archaic population, $D$. (The outgroup is omitted from this figure for
simplicity.) As the figure illustrates, \emph{both} forms of archaic
admixture can inflate the count of site pattern $ny$. For this reason,
ghost admixture may bias our estimate of $m_N$.

To describe these effects, we introduce a new notation.  Let $I_{ny}$
represent the number of sites exhibiting pattern $ny$, i.e.\ the
number at which the derived allele is carried only by samples $n$ and
$y$. We assume for the moment that we have no sample from the second
archaic population, $D$. It contributes DNA to our sample only via
gene flow into population $Y$. Thus, we have three $I$ statistics:
$I_{ny}$, $I_{nx}$, and $I_{xy}$. These definitions are summarized in
table~\ref{tab.pat}.  If either archaic population contributed genes
to population $Y$, then $I_{ny}$ should exceed $I_{nx}$. This is
indeed the case for the data summarized in table~\ref{tab.pat}. Our
goal is to use such discrepancies to study the rates, $m_N$ and $m_D$,
of gene flow from archaics into population $Y$.

\subsection{Expectations of ratios}
\label{sec.Eratio}

All the estimators discussed here use a ratio of expectations to
approximate the expectation of a ratio.  In each case, numerator and
denominator are sums across all polymorphic sites in the sample.  If
the number of such sites is large, the weak law of large numbers
implies that numerator and denominator should each be close to their
expectations, so that the ratio of expectations is a good
approximation.  This approximation is quite accurate when samples are
entire genomes. It would be less accurate in smaller regions, such as
individual genes.

\begin{table*}
\caption{Symbols and reference values for time parameters}
\label{tab.time}
{\centering
\begin{tabular}{crp{0.4\textwidth}p{0.3\textwidth}}
Symbol & Value & Definition & Reference\\
       & (ky)\\
\hline\hline
$\alpha$
  & 25
  & latest plausible Denisovan admixture
  & a guess\\
$\alpha'$ 
  & 50
  & earliest plausible Denisovan admixture
  & a guess\\
$\beta$  
  & 40
  & separation time of Melanesians and other Eurasian populations
  \\
$\iota$
  & 39.7
  & age of Mezmaiskaya fossil
  & \citep{Pinhasi:PNA-108-8611}\\
$\gamma$
  & 40
  & age of Denisovan fossil
  & \citep[p.~1053]{Reich:N-468-1053}\\
$\delta$ 
  & 55 
  & time of Neanderthal admixture
  & \citep[table~2]{Sankararaman:PLO-8-e1002947}\\
$\varepsilon$ 
  & 65 
  & age of older Neanderthal fossil used in $\hat f$ or of Altai Neanderthal
  & \citetext{\citealp[pp.~713]{Green:S-328-710}; \citealp[table~1]{Prufer:N-505-43}}\\
$\theta$
  & 95
  & separation time of populations of Mezmaskaya and Altai
  Neanderthals
  & \citep[p.~44]{Prufer:N-505-43}\\
$\zeta$  
  & 110 
  & separation time of Africans and Eurasians
  & \citep[p.~153]{Veeramah:NRG-15-149}\\
$\eta$   
  & 200
  & separation time of San and other modern human populations\\
$\kappa$ 
  & 427 
  & separation time of Neanderthal and Denisovan populations
  & \citep[p.~44]{Prufer:N-505-43}\\
$\lambda$
  & 658 
  & separation time of modern and archaic populations
  & \citep[p.~44]{Prufer:N-505-43}\\
$\mu$
  & 
  & separation time of chimpanzee and hominin populations\\
\hline
\end{tabular}\\}
\end{table*}

\subsection{Time parameters}
\label{sec.timeval}

Greek letters represent time parameters, as summarized in
table~\ref{tab.time}. The values there are in kiloyears (ky).  Where
possible, these are estimates taken from the literature, but we have
not included statistical uncertainties because of our focus in this
article on expected values and bias. Where published values are
ranges, we use the midpoints.  The parameters $\alpha$ and $\alpha'$
are intended bracket the time of gene flow from Denisovans into
Eurasian populations. Although there are no compelling estimates of
these parameters, it seems unlikely that this gene flow could have
occurred much before 50~ky ago or much after 25~ky ago.

In calculations, the values in table~\ref{tab.time} are re-expressed
in units of $2N_0$ generations, where the generation time is 25 years,
and $N_0=2308$ is the estimated size of the ancestral human
population, as shown in table~\ref{tab.psize}.

\subsection{Simulations to validate theory}
\label{sec.validation}

To check for errors in algebraic formulas, we developed software to
simulate the components of each estimator. These simulations generate
gene genealogies by running a coalescent process constrained by
assumptions involving the population tree (branch lengths as well as
topology), admixture events (timing as well as level of gene flow),
and the size of each population. For details, see
section~\ref{S-sec.validation} of \emph{Supplementary Materials}. 

\section{Results}
\label{sec.results}

\subsection{The estimator $\hat f$ of \citet{Patterson:S-328-S158}}  
\label{sec.results-fhat}

The first estimator to use the idea outlined in
section~\ref{sec.logic} was that of
\citet[Eqn.~S18.5]{Patterson:S-328-S158}. Their estimator
$\hat f$ is designed to estimate Neanderthal gene flow. It can be 
written as
\begin{equation}
\hat f = \frac{I_{ny} - I_{nx}}{J_{n'n}-J_{nx}}
\label{eq.f}
\end{equation}
where $n$ and $n'$ are two genomes sampled from the Neanderthal
population $N$. The numerator and denominator analyze different sets
of genomes: $\{n, x, y\}$ and $\{n, n', x\}$. Because these sets
differ, the number of sites with pattern $nx$ will not in general be
identical in the numerator and denominator. For this reason, we use
$J$s rather than $I$s to represent counts of sites in the denominator.

In practice, \citet[p.~159]{Patterson:S-328-S158} use a weighted
average of two such estimators, in which the roles of the two
Neanderthal fossils are switched. This does not change expected
values, so we use the simpler formulation in Eqn.~\ref{eq.f}.

\ref{sec.f} derives the expected value of $\hat f$ under a generalized
model (Fig.~\ref{fig.xynd}), which allows for ghost admixture from
archaic population $D$. The expected value is
\begin{subequations}
\begin{align}
E[\hat f] &\approx
\frac{m_N(1-m_D) (\lambda-\delta + s_1) + m_D(\lambda-\kappa  + s_2)}%
{(1-m'_D) (\lambda-\varepsilon + s_3) + m'_D(\lambda-\kappa  + s_2)}
\label{eq.Ef1}\\
&\approx
  m_N \left(\frac{\lambda-\delta + s_1}{\lambda-\varepsilon + s_3}\right)
+ m_D \left(\frac{\lambda-\kappa + s_2}{\lambda-\varepsilon +
  s_3}\right)
\label{eq.Ef}
\end{align}
\end{subequations}
where $m_D$ and $m'_D$ are rates of gene flow from $D$ into $Y$ and
into $N$, and (\ref{eq.Ef}) ignores 2nd-order terms in $m_N$, $m_D$,
and $m'_D$.  The $s_i$ terms are
\begin{align*}
s_1 &= (1-K_N)F_N^{(\delta,\kappa)} + (1-K_{ND})S_N^{(\delta,\kappa)}F_{ND}^{(\kappa,\lambda)}\\
s_2 &= (1-K_{ND})F_{ND}^{(\kappa,\lambda)}\\
s_3 &= (1-K_N)F_N^{(\delta,\kappa)} +
(1-K_{ND})S_N^{(\varepsilon,\kappa)}F_{ND}^{(\kappa,\lambda)}
\end{align*}
and measure population-size bias. If all populations are of equal
size, $s_i=0$ for all $i$.

The expression $(1-K_N)F_N^{(\delta,\kappa)}$, which appears
in $s_1$ and $s_3$, accounts for any difference in size
between the ancestral human population and \textit{N}, the Neanderthal
population. If \textit{N} were small, a pair lineages would tend to
coalesce within it, so $F_N^{(\delta,\kappa)}$ would be large. In
addition, the early coalescence would lengthen the branch along which
mutations can generate site pattern $ny$.  The factor $1-K_N$ accounts
for this change in branch length, as explained in \ref{sec.wi}. Taken
together, these effects make $s_1$ and $s_3$ decreasing functions of
$K_N$.

Eqn.~\ref{eq.Ef} shows that $\hat f$ estimates $m_N$, but not without
bias. The bias disappears only if ghost admixture is absent ($m_D=0$),
population sizes are equal ($s_i=0$), and the older fossil lived at
the time of the gene flow from $N$ into $Y$ ($\delta=\varepsilon$). If
the first two of these conditions hold, Eqn.~\ref{eq.Ef} is equivalent
to the result of \citet[p.~55]{Patterson:N-468-S47}. We consider the
magnitudes of these biases below.

\subsection{The estimator $R_N$ of \citet{Patterson:N-468-S47}}
\label{sec.results-RN}

Another method was needed for studying Denisovan admixture, because
only one Denisovan genome is available.  \citet{Patterson:N-468-S47}
introduce two new statistics. The first of these,
$R_{\text{Neanderthal}}$ (their Eqn.~S8.3), estimates gene flow from
Neanderthals.  \citet[p.~42]{Patterson:S-338-222-36} call this
statistic ``Nea.''  They also define an analogous statistic, ``Den,''
in which the roles of Neanderthal and Denisovan samples are reversed,
and the goal is to estimate Denisovan admixture. The analysis for
these statistics is the same, assuming that the Neanderthal and
Denisovan populations are sister taxa within the population tree.

\begin{figure*}
{\centering\input{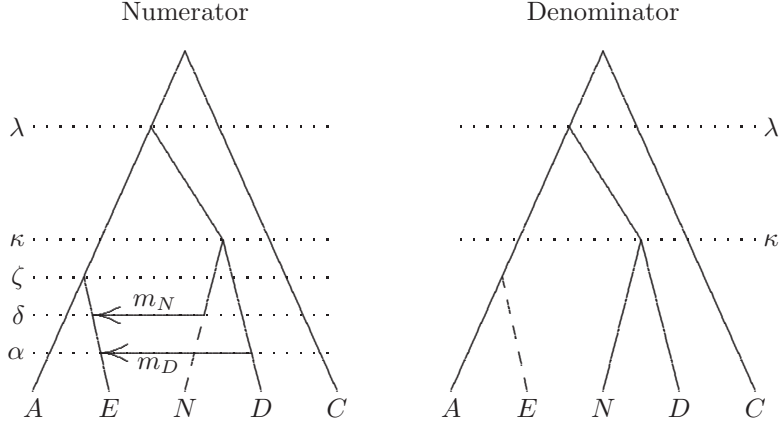}\\}
\caption{Population tree for $R_N$
  \citep[Eqn.~S8.3]{Patterson:N-468-S47}. The populations involved are
  Africa $(A)$, Eurasia $(E)$, Neanderthal $(N)$, Denisova $(D)$, and
  chimpanzee $(C)$. Arrows indicate gene flow from Neanderthals at rate
  $m_N$ and, later, from Denisovans at rate $m_D$. Only four
  populations are compared in the numerator of $R_N$, and a different
  four are compared in the denominator. In each panel of the figure,
  dashed line indicates the sample being ignored. Greek letters
  represent the times of various events.}
\label{fig.RNtree}
\end{figure*}

We abbreviate $R_{\text{Neanderthal}}$ as $R_N$ and define it as
\begin{equation}
R_N = \frac{I_{de} - I_{ad}}{J_{dn} - J_{ad}}
\label{eq.RN}
\end{equation}
In this statistic, the numerator compares an African genome, $a$, a
Eurasian genome, $e$, and the Denisovan genome, $d$. A chimpanzee
genome is used to infer the ancestral state. The denominator drops the
Eurasian genome and adds a Neanderthal genome, $n$. These genomes are
sampled from populations $A$, $E$, $N$, $D$, and $C$, using the model
in Fig.~\ref{fig.RNtree}.

We consider a model in which the Eurasian population received gene
flow first from Neanderthals, at time $\delta$, and then from
Denisovans, at time $\alpha$. The first of these episodes replaced a
fraction $m_N$ of the Eurasian gene pool; the second replaced a
fraction $m_D$. The goal is to estimate Neanderthal gene flow, so
$m_D$ measures ghost admixture.

\begin{figure}
{\centering\input{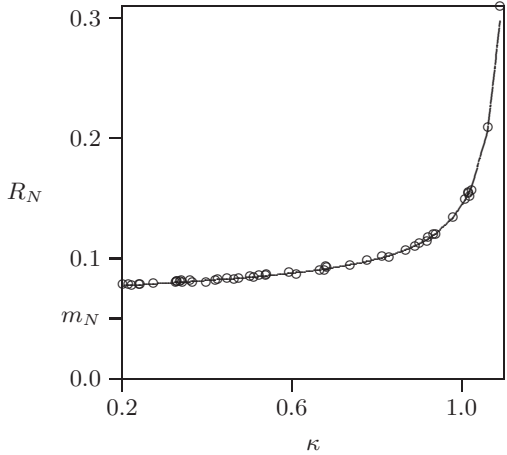}\\}
\caption{Response of $R_N$ to the separation time, $\kappa$, of
  Neanderthals and Denisovans. Circles show simulated values and solid
  line shows theoretical formula (Eqn.~\ref{eq.ERN}). Simulations
  generate branch lengths only (not mutations) and are described in
  section~\ref{sec.validation}. They use $10^6$ iterations, assume
  $m_N=0.05$, $m_D=0.025$, and other parameters as in
  tables~\ref{tab.psize}--\ref{tab.time}. $\kappa$ is in units of
  $2N_0$ generations.}
\label{fig.RNkappa}
\end{figure}

\ref{sec.RN} derives the expectation of $R_N$,
\begin{equation}
E[R_N] \approx m_N(1-m_D)
 + m_D\left(\frac{\lambda-\alpha +s_4}{\lambda-\kappa + s_5}\right)
\label{eq.ERN}
\end{equation}
The $s_i$ terms are
\begin{align*}
s_4 &= (1-K_D)F_D^{(\alpha,\kappa)} 
 + (1-K_{ND})S_D^{(\alpha,\kappa)}F_{ND}^{(\kappa,\lambda)}\\
s_5 &= (1-K_{ND})F_{ND}^{(\kappa,\lambda)}
\end{align*}
and measure population-size bias.  If ghost admixture is absent
($m_D=0$), then $R_N$ provides an unbiased estimate of $m_N$
\citetext{\citealp[Eqn.~S8.4]{Patterson:N-468-S47};
  \citealp[S11.9]{Durand:N-468-S169}}. Otherwise, it is affected by
all three forms of bias. Fig.~\ref{fig.RNkappa} shows how it varies in
response to $\kappa$, the Neanderthal-Denisovan separation time.

\subsection{The estimator $R_D$ of \citet{Patterson:N-468-S47}}
\label{sec.results-RD}

\begin{figure*}
{\centering\input{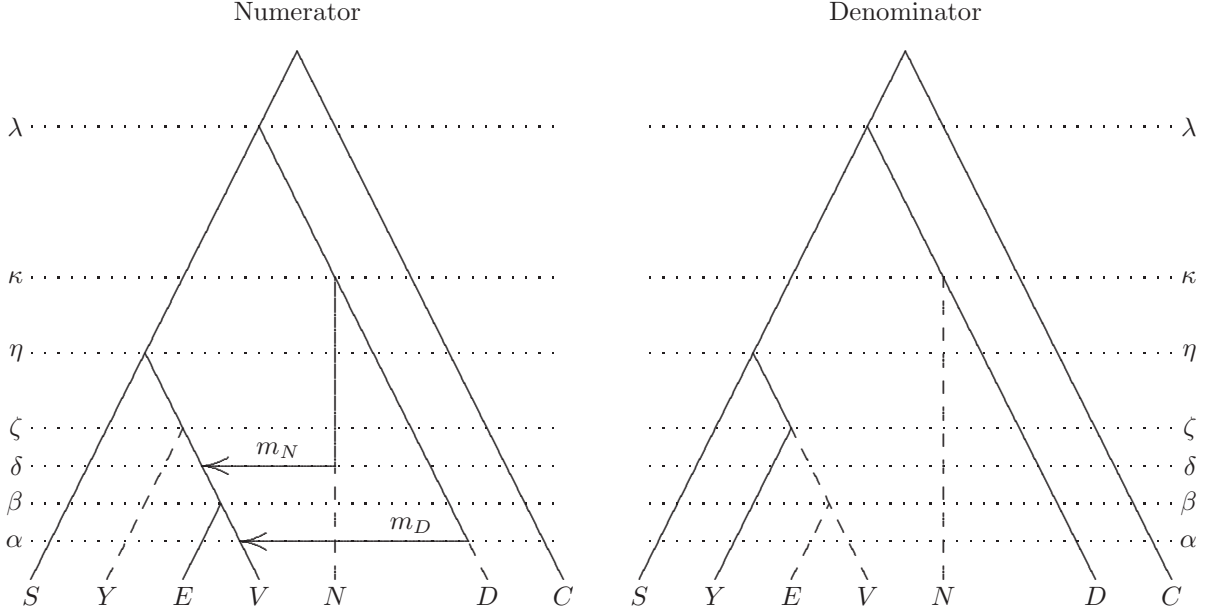}\\}
\caption{Population tree for $R_D$
  \citep[Eqn.~S8.5]{Patterson:N-468-S47}. The populations involved are
San $(S)$, Yoruba $(Y)$, Eurasia $(E)$, Melanesia $(V)$, Neanderthal
$(N)$, Denisova $(D)$, and chimpanzee $(C)$. Arrows indicate gene flow
from Neanderthals, at rate $m_N$, and Denisovans, at rate
$m_D$. Although seven populations are involved, only four are compared
in the numerator of $R_D$, and a different four are compared in the
denominator. In each panel of the figure, dashed lines indicate the
samples being ignored. Greek letters represent the times
of various events.}
\label{fig.RDtree}
\end{figure*}

\citet[Eqn.~S8.5]{Patterson:N-468-S47} also define a second statistic,
$R_{\text{Denisova}}$, which we abbreviate as
$R_D$. It can be written as
\begin{equation}
R_D = \frac{I_{es} - I_{sv}}{J_{sy} - J_{ds}}
\label{eq.RD}
\end{equation}
The numerator compares genomes sampled from three populations: San,
$S$, Eurasian, $E$, and Melanesian, $V$. The chimpanzee, $C$,
determines ancestral state. The denominator drops $E$ and $V$ but adds
Yoruban, $Y$, and Denisovan, $D$. These populations are related as in
Fig.~\ref{fig.RDtree}. This model assumes that the ancestral Eurasian
population exchanged a fraction $m_N$ of its genes with
Neanderthals. Later, the Melanesian population exchanged a fraction
$m_D$ with Denisovans. The goal is to estimate $m_D$, so $m_N$
measures ghost admixture.

\ref{sec.RD} derives an approximation to the expected value of $R_D$,
\begin{equation}
E[R_D] \approx m_D(1-m_N)
\label{eq.ERD}
\end{equation}
in agreement with published results
\citetext{\citealp[Eqn.~S8.6]{Patterson:N-468-S47};
  \citealp[S11.19]{Durand:N-468-S169}}.  Of the estimators considered
here, this is the only one unaffected by bias involving branch lengths
and changes in population size. It is sensitive to ghost admixture
($m_N$), but only slightly unless $m_N$ is large.  In an effort to
remove this bias, \citet[p.~57]{Patterson:N-468-S47} estimate $m_D$ as
$R_D/(1-R_N)$. This involves using $R_N$ with a population that
received Denisovan gene flow. Because $R_N$ is biased in such cases,
this procedure imports bias into the estimate of $m_D$. This effect
will be small, however, unless $R_N$ is large.

\subsection{The estimator $p_D$ of \citet{Reich:AJH-89-516}}
\label{sec.results-pD}

\begin{figure}
{\centering\input{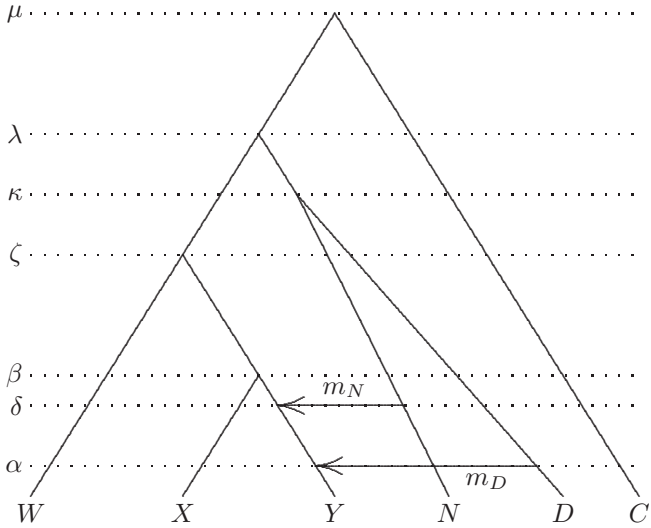}\\}
\caption{Population tree for $p_D$. $W$, $X$, and $Y$ are three modern
  human populations; $N$ and $D$ are two archaic populations, and $C$
  is the chimpanzee population. A fraction $m_D$ of the gene pool of
  $Y$ descends via gene flow from $D$. Of the remaining fraction
  $1-m_D$, a fraction $m_N$ descends via gene flow from $N$. Greek
  letters represent the times of various events.}
\label{fig.pD}
\end{figure}

\citet[p.~1072]{Patterson:G-192-1065} define a family of related
statistics, which they call ``$F_4$-ratio estimators.''
\citet[Eqn.~1]{Reich:AJH-89-516} use one of these to study how
Denisovan admixture varies from population to population:
\begin{equation}
p_D = \frac{\sum_i (w_i - d_i)(x_i - y_i)}{\sum_i (w_i - d_i)(x_i -
  y'_i)}
\label{eq.pDdef}
\end{equation}
In this equation, lower-case letters do not represent individual
genomes. Instead, $w_i$, $x_i$, $y_i$, and $d_i$ are frequencies of a
given allele at locus $i$ within four populations, $W$, $X$, $Y$, and
$D$. In the denominator, $y'_i$ is the frequency within a fifth
population, $Y'$. The model assumes that $Y'$ occupies a position
within the population tree that is similar to that of $Y$. In other
words, the tree of $W$, $X$, $Y'$, and $D$ is identical to that in
figure~\ref{fig.pD}. The goal is to estimate the ratio $m_D/m'_D$ of
Denisovans admixture into $Y$ and $Y'$. The corresponding Neanderthal
rates, $m_N$ and $m'_N$, measure ghost admixture.

\ref{sec.pD} derives the expected value of $p_D$ under a mutational
clock. This is a departure from \citet{Reich:AJH-89-516}, who couch
their theory in terms of genetic drift. Although Reich et al.\ define
$w_i$, $x_i$, $y_i$, and $d_i$ as frequencies of an arbitrary allele,
these become derived allele frequencies in our analysis. We drop the
assumption \citep[p.~1069]{Patterson:G-192-1065} that the locus is
polymorphic at the root of the human tree. The expected value of $p_D$
is approximately
\begin{equation}
E[p_D] \approx \frac{m_D A + m_N(1-m_D) B}{m'_D A'  + m'_N(1-m'_D)B}
\label{eq.EpD}
\end{equation}
where 
\begin{align*}
A &= 2\lambda-\zeta -\alpha + s_6\\
A' &= 2\lambda-\zeta -\alpha' + s_7\\
B &= 2\lambda - \zeta - \kappa + s_8
\end{align*}
and $\alpha$ and $\alpha'$ are the times of Denisovan gene flow into
populations $Y$ and $Y'$. The $s_i$,
\begin{align*}
s_6 &= 2T_0^{(\lambda)} - T_{WXY}^{(\zeta)} - T_D^{(\alpha)}\\
s_7 &= 2T_0^{(\lambda)} - T_{WXY}^{(\zeta)} - T_D^{(\alpha')}\\
s_8 &= 2T_0^{(\lambda)} - T_{WXY}^{(\zeta)} - T_{ND}^{(\kappa)}
\end{align*}
measure population-size bias and disappear under constant size.

If population size is constant, Neanderthal admixture is absent, and
the two Denisovan admixtures were simultaneous, then our results reduce
to $E[p_D] = m_D/m'_D$, in agreement with
\citet[p.~523]{Reich:AJH-89-516}.

\subsection{The estimator $p_N$}
\label{sec.results-pN}

\begin{figure}
{\centering\input{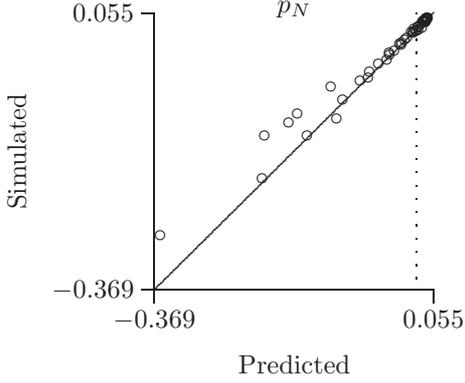}\\}
\caption{Population tree for $p_N$. $A$ is the modern African
  population, $X$ is a modern admixed population, $I$ is the
  introgressing Neanderthal population, $M$ is the population of the
  Mezmaiskaya Neanderthal, $N$ that of the Altai Neanderthal, $D$ is
  Denisova, and $C$ is chimpanzee. A fraction $m_D$ of the gene pool of
  $X$ descends via gene flow from $D$. Of the remaining fraction
  $1-m_D$, a fraction $m_N$ descends via gene flow from $I$. Greek
  letters represent the times of various events.}
\label{fig.pN}
\end{figure}

\citet[Eqn.~S14.13]{Patterson:N-505-S120} use another $F_4$-ratio
statistic to estimate the level of Neanderthal admixture into
Europeans. Although they refer to it as $\hat\alpha$, we refer to it
as $p_N$ to avoid confusion with the time parameter, $\alpha$, defined
in our table~\ref{tab.time}. It is defined as
\begin{equation}
p_N = \frac{\sum_i (d_i - n_i)(a_i - x_i)}{\sum_i (d_i - n_i)(a_i -
  m_i)}
\label{eq.pNdef}
\end{equation}
As with $p_D$, the lower-case letters represent not haploid genomes,
but allele frequencies within four populations, which are related as
shown in Fig.~\ref{fig.pN}. We generalize the model of Patterson et
al.\ to allow for ghost admixture (from Denisovans) as well as primary
admixture (from Neanderthals).

We analyze this statistic under a molecular clock.  The expectation of
$p_N$, as derived in \ref{sec.pN}, is
\begin{equation}
E[p_N] \approx m_N(1-m_D)
- m_D\left(\frac{\kappa-\alpha+s_9}{\kappa-\theta+ s_{10}}\right)
\label{eq.EpN}
\end{equation}
where the terms
\begin{align*}
s_9 &= T_{IMND}^{(\kappa)}-T_D^{(\alpha)}, \quad\text{and}\\
s_{10} &= T_{IMND}^{(\kappa)} - T_{IMN}^{(\theta)}
\end{align*}
measure population-size bias.  In the absence of ghost admixture,
$m_D=0$, and $E[p_N] \approx m_N$, in agreement with
\citet[Eqn.~S14.2]{Patterson:N-505-S120}. In this case, $p_N$
is unbiased.

When ghost admixture is present, however, substantial biases arise.
The interval ($\kappa-\alpha$) in the numerator extends from the
Neanderthal-Denisovan separation to the admixture of Denisovans into
moderns. Although this interval is not well constrained, it must have
been hundreds of thousands of years. The interval ($\kappa-\theta$) in
the denominator extends from the Neanderthal-Denisovan separation to
that of the populations of the Mezmaskya and Altai Neanderthals and is
probably much shorter.  This suggests that, in Eqn.~\ref{eq.EpN},
the coefficient of $m_D$ may exceed that of $m_N$. If so, $p_N$
is more sensitive to ghost admixture than to primary admixture.

The coefficient of $m_D$ also includes terms involving changes in
population size. Although $p_N$ is unbiased in the absence of
ghost admixture, it is affected by all three forms of bias when ghost
admixture is present.

\subsection{A new estimator}
\label{sec.results-Q}
Let us revisit the history illustrated in Fig.~\ref{fig.xynd}, whose
properties are studied in \ref{sec.f}. We propose to estimate $m_N$
with the ratio
\begin{equation}
Q = \frac{I_{ny} - I_{nx}}{I_{xy}-I_{nx}}
\label{eq.Q}
\end{equation}
Unlike most of those published previously, this statistic requires
only one archaic genome.  The expectation of $Q$ is approximately the
ratio of Eqns.~\ref{eq.Inydiff} and~\ref{eq.Ixydiff},
\begin{subequations}
\label{eq.EQ}
\begin{align}
E[Q] &\approx
\frac{m_N(1-m_D) (\lambda-\delta + s_1) + m_D(\lambda-\kappa  + s_2)}%
{(1-m_N)(1-m_D)(\lambda-\zeta + s_{11}\bigr)}
\label{eq.EQ1}\\
&\approx 
m_N\left(\frac{\lambda-\delta + s_1}{\lambda-\zeta + s_{11}}\right)
+ m_D\left(\frac{\lambda-\kappa + s_2}{\lambda-\zeta + s_{11}}\right)
\label{eq.EQapprx}
\end{align}
\end{subequations}
where $s_1$ and $s_2$ are as defined in
section~\ref{sec.results-fhat}, and
\[
s_{11} = (1-K_{XY}) F_{XY}^{(\zeta,\lambda)}
\]
$Q$ is sensitive to all three forms of bias but is especially
sensitive to population-size differences. Recent estimates indicate
that archaic populations were much smaller than early modern ones
\citep[Fig.~4]{Prufer:N-505-43}. This makes $s_1$ and $s_2$ larger
than $s_{11}$ and biases $Q$ upwards. Branch lengths provide an
additional upward bias, because $\lambda-\delta$ probably exceeds
$\lambda-\zeta$.

\subsection{Effects of branch lengths and population sizes}
\label{sec.results-timesize}

To get a sense of magnitudes, let us calculate the expected value of
each statistic under standard assumptions. We use the parameter values
in tables~\ref{tab.psize}--\ref{tab.time}, take the level of primary
admixture as 0.05 and that of secondary admixture as 0.025. For $p_D$,
we set $m_D=0.1$ in the numerator and $m'_D=0.05$ in the denominator.
With these values,
\begin{equation}
\renewcommand{\arraystretch}{1.1}
\left.\begin{array}{rcr@{\;\times\;}l}
E[\hat f] &\approx& 1.24 & m_N\\
E[R_N]    &\approx& 2.11 & m_N \\
E[R_D]    &\approx& 0.98 & m_D\\
E[p_D]    &\approx& 0.89 & m_D/m'_D\\
E[p_N]&\approx& 0.37 & m_N\\
E[Q]      &\approx& 1.92 & m_N/(1-m_N)
\end{array}\right\}
\label{eq.bias}
\end{equation}
These should not be viewed as precise numerical estimates of
bias. Instead, they are representative values under one set of
parameter values. Under these assumptions, the
estimators differ greatly in bias. The multiplier for $R_D$ is close
to unity, indicating that it is nearly unbiased. The other estimators
have substantial biases.

These biases could be corrected, given branch lengths, population
sizes and the rate of ghost admixture, by equating observed to
expected values and solving for primary admixture. The trouble is that
these parameters cannot be estimated precisely, and any error would
bleed into estimates of admixture. To measure sensitivity to these
uncertainties, we calculate \emph{elasticities}.

\begin{table}
\setlength{\tabcolsep}{5pt}%
\caption{Elasticities of $\hat f$, $R_N$, $p_D$, $p_N$, and $Q$
  with respect to times and population sizes. The table assumes a
  level, 0.05, of primary admixture and 0.025 of ghost admixture, and
  other parameters as in tables~\ref{tab.psize}--\ref{tab.time}. For
  $p_D$, we assume that $m_D=0.1$ and $m'_D=0.05$. Elasticities with
  absolute values greater than 0.5 are in bold type.}
\label{tab.elasticity}
{\centering
\begin{tabular}{lrrrrr}
\hline
           & $\hat f$ & $R_N$ & $p_D$ & $p_N$ & $Q$\\
\cline{2-6}
$\lambda$  &   0.187
           & \textbf{--0.623}
           & --0.081
           &
           & --0.247\\
$\kappa$   & --0.241
           & \textbf{0.724}
           & 0.063
           & 0.364
           &  --0.257\\
$\delta$   &  --0.065
           &
           & 
           & 
           & --0.065\\
$\theta$   &
           &
           & 
           & --0.438
           &\\
$\varepsilon$&   0.095
           &
           &      
           &
           & \\
$\zeta$    &  
           &
           & 0.006
           &
           & 0.237\\
$\alpha$   &  
           & --0.019
           & 
           & --0.101
           & \\
$\alpha'$  &  
           &
           & 0.033
           &
           & \\
$K_N$      & 0.008
           &
           & 
           &
           & --0.034\\
$K_D$      &
           & --0.022
           & 0.004
           & 0.093
           & \\
$K_{ND}$    & --0.016
           & 0.049
           & 0.004
           &
           & --0.017\\
$K_{IMND}$  & 
           & 
           & 
           & --0.006
           & \\
$K_{IMN}$  & 
           & 
           & 
           & --0.097
           & \\
$K_{XY}$    &
           &
           & 
           &
           & 0.415\\
$K_{WXY}$    &
           &
           & 0.054
           &
           & \\
\hline
\end{tabular}\\}
\end{table}

Elasticity is the proportional change in one quantity caused by a
given proportional change in another.  For example, the elasticity of
$\hat f$ with respect to $\lambda$ is $d \log E[\hat f]/d \log
\lambda$. $R_D$ has elasticity zero with respect to branch lengths and
population sizes. The elasticities of the other statistics are shown
in table~\ref{tab.elasticity}, evaluated at the values in
tables~\ref{tab.psize}--\ref{tab.time}.  Elasticities larger than 0.5
are shown in bold. The elasticity with largest absolute value, 0.724,
is that of $R_N$ with respect to $\kappa$. If the true value of
$\kappa$ were 10\% larger than our estimate, $R_N$ would be inflated
by about 7\%. Fig.~\ref{fig.RNkappa} shows this response in greater
detail. 

The relatively large elasticities of $R_N$, $p_N$, and $Q$ make
them sensitive to error in parameter estimates. The outlook is
brighter for $\hat f$, $p_D$, and $R_D$, which have smaller
elasticities.

\subsection{Effect of ghost admixture}
\label{sec.bias}

Each statistic is designed to estimate admixture from a specific
archaic population. We refer to this as ``primary admixture.''
However, the expected values of these statistics also depend on
``ghost'' admixture from other archaic populations.

\begin{figure}
{\centering\input{figfrance}\\}
\caption{Archaic admixture in France, as implied by estimates of $\hat
  f$ \citep[table~S58]{Patterson:S-328-S158}, $Q$ (this article),
  $R_N$ \citep[p.~55]{Patterson:N-468-S47}, and $p_N$
  \citep[p.~128]{Patterson:N-505-S120}. Curves assume values in
  tables~\ref{tab.psize}--\ref{tab.time}.}
\label{fig.france}
\end{figure}

To see this effect, consider the statistic $Q$. Subsititing from
table~\ref{tab.pat} into Eqn.~\ref{eq.Q} gives
\[
Q = \frac{103,612 - 95,347}{303,340 - 95,347} \approx 0.0397 
\]
After setting $Q=E[Q]$, Eqn.~\ref{eq.EQ} defines $m_N$ as an implicit
function of $m_D$. This function is shown as a dark blue dashed line
in Fig.~\ref{fig.france}. The slope of that line measures the
sensitivity of $Q$ to ghost admixture. If ghost admixture is absent,
this estimate suggests that a fraction $m_N = 0.025$ of French DNA
derives from Neanderthals. This estimate does not equal $Q$, because
it corrects for bias using the parameter values in
tables~\ref{tab.psize}--\ref{tab.time}. If ghost admixture were 5\%,
on the other hand, we would conclude that $m_N=0$. By itself, $Q$
cannot choose between these alternatives.

This ambuguity evaporates when we consider other estimators. The other
curves in Fig.~\ref{fig.france} refer to other statistics that have
been used to estimate Neanderthal admixture into the French
population. The curves have very different slopes, because the
estimators respond differently to ghost admixture. In the absence of
sampling error and uncertainty about parameter values, they should all
intersect at a point corresponding to the true values of $m_N$ and
$m_D$. Although the curves do not intersect at a point, the difference
between them is smallest near the left edge of the graph, where $m_D$
is small. This suggests that the French received archaic gene flow
primarily from Neanderthals, in agreement with
\citet[p.~1056]{Reich:N-468-1053}.

The agreement between $Q$ and the other statistics also tells us
something about ancient population sizes. As
table~\ref{tab.elasticity} shows, $Q$ is sensitive to population
sizes. Indeed, if one builds a graph like Fig.~\ref{fig.france}
\emph{without} correcting for population size, $Q$ is not consistent
with the other estimators (data not shown). In correcting for
population size, we have assumed that archaic populations were much
smaller than early modern ones, as shown in table~\ref{tab.psize}. Had
this assumption been incorrect, the correction would not have
helped. Thus, the consistency between $Q$ and the other estimators
supports current views about the sizes of ancient populations
\citetext{\citealp[Fig.~5]{Meyer:S-338-222};
  \citealp[Fig.~4]{Prufer:N-505-43}}.

\begin{figure}
{\centering\input{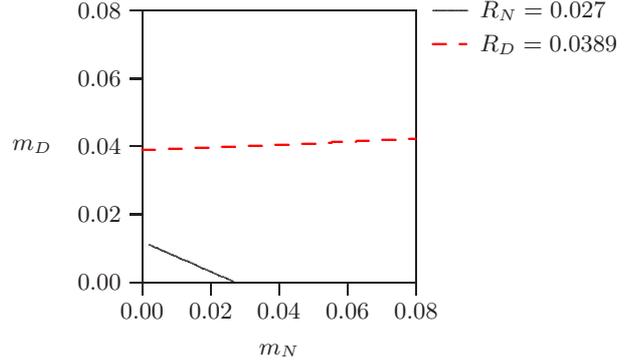}\\}
\caption{Archaic admixture in Melanesia, as implied by the observation
  that $R_N=0.027$ and $R_D/(1-R_N) = 0.04$
  \citep[p.~58]{Patterson:N-468-S47}. Curves assume values in
  tables~\ref{tab.psize}--\ref{tab.time}.}
\label{fig.melanesia}
\end{figure}

Fig.~\ref{fig.melanesia} attempts a similar analysis with Melanesian
data. For that population, \citet[p.~58]{Patterson:N-468-S47} estimate
that $R_N = 0.027$ and $R_D/(1-R_N) = 0.04$, which implies that $R_D =
0.0389$. At face value, these data suggest that the Melanesian gene
pool includes a small contribution from Neanderthals and a larger one
from Denisovans. The curves in Fig.~\ref{fig.melanesia}, however, do
not support this view. The two curves are far apart throughout the
horizontal range. Even if Melanesians received no Neanderthal
admixture at all, $R_N$ should be much higher than the observed value.

There are several conceivable explanations for this
discrepancy. First, it could result from sampling error in the
estimates of $R_N$ and $R_D$. This seems unlikely, in view of the
large sample of sites, but cannot be excluded without statistical
analysis. We do not attempt such an analysis here. Second, the
discrepancy could result from error in the parameter values in
tables~\ref{tab.psize}--\ref{tab.time}. It would be useful to fit
these parameters rather than taking their values as given. Finally,
the discrepancy could arise because the model is
misspecified---because the history of the populations under study
violates the assumptions in Figs.~\ref{fig.RNtree} and
\ref{fig.RDtree}. We discuss this last possiblity below.

\section{Discussion and Conclusions}
\label{sec.discuss}

This article has explored biases that arise in one family of
estimators---those based on the frequency with which derived alleles
are shared by pairs of samples. We have focused on three forms of
bias, involving branch lengths in the population tree, differences in
population size, and ghost admixture.

All of these estimators are sensitive to ``ghost admixture''---to
admixture from archaic populations other than the one of primary
interest. The most robust estimator, $R_D$, is hardly affected.  The
others are at least modestly sensitive. For $R_N$ and $p_N$,
this effect is profound.

Branch-length bias arises because the effect of admixture depends not
only on the number of immigrants, but also on the genetic difference
between immigrants and residents. The longer the two populations have
been separated, the greater their genetic difference, and the greater
the effect of any given level of gene flow. For this reason, the
expected values of admixture statistics may depend on branch lengths.

Yet in the absence of ghost admixture, most published estimators are
relatively free of branch-length bias. Each is constructed as a
ratio. When branch length effects are identical in numerator and
denominator, they cancel, and the resulting expression is free of
branch-length bias. This cancellation often fails, however, in the
presence of ghost admixture. In that case, there may be two
branch-length effects, which cannot both cancel in a single ratio. For
this reason, branch lengths bias $R_N$ and $p_N$ only in the
presence of ghost admixture. With other estimators, branch lengths
contribute bias even in the absence of ghost admixture. Only $R_D$
escapes this effect.

Where branch-length effects exist, they are accompanied by terms
involving differences in population size. These differences distort
branch lengths within the gene genealogy and thus alter the
probability that mutations will generate particular site
patterns. These effects, however, are generally small, at least for
the parameter values in tables~\ref{tab.psize}--\ref{tab.time}.
Population-size bias is substantial only for the new estimator, $Q$.

The estimators respond to ghost admixture in different ways, as shown
by their differing slopes in Figs.~\ref{fig.france}
and~\ref{fig.melanesia}. Because of these differences, comparisons
among estimators provide information. Such comparisons indicate that
archaic gene flow into Europe came primarily from Neanderthals and
support the view that archaic populations were much smaller than those
of early modern humans. They also expose an inconsistency in estimates
of archaic admixture into Melanesia. As shown in
Fig.~\ref{fig.melanesia}, no pair of $(m_N, m_D)$ values is consistent
both with the observed value of $R_N$ and also with that of $R_D$.

Although this inconsistency may be a statistical artifact, it could
also result from an incorrectly specified model. For example,
\emph{Homo erectus} may have contributed genes to populations of
Denisovans \citep[p.~48]{Prufer:N-505-43} or to modern humans in
Melanesia \citep{Mendez:MBE-29-1513}. Such gene flow would violate the
assumptions underlying our analysis of $R_N$ and $R_D$ and might
account for the discrepancy seen in Fig.~\ref{fig.melanesia}.

Our empirical conclusions should be regarded with caution, because we
have made no effort to account for statistical uncertainty.
Nonetheless, the exercise illustrates that the biases in these
estimators are not altogether bad. Because these estimators respond to
population history in different ways, comparisons among them provide 
new information.

\appendix

\section{Probabilities of site patterns}
\label{sec.sitepat}

\begin{figure*}
{\centering\input{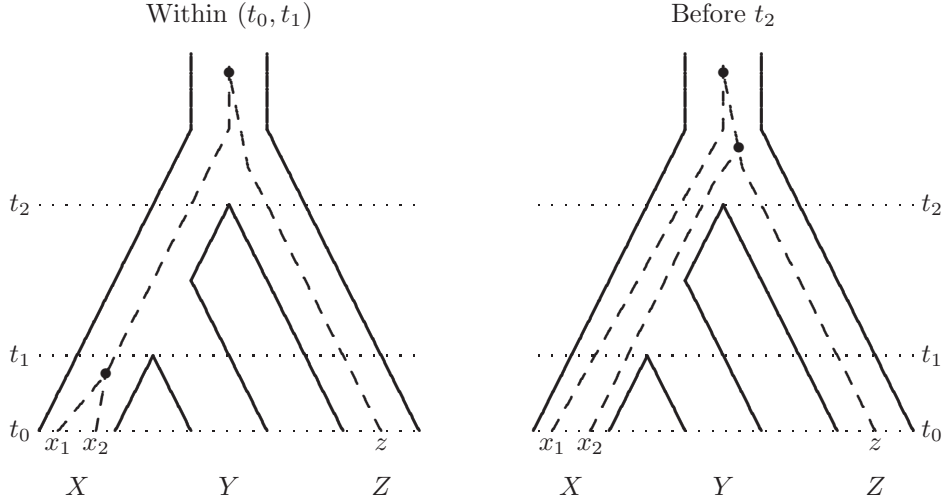}\\}
\caption{Coalescent events within an interval (left panel) and in the
  ancestral population (right panel). During interval $(t_0,t_1)$, a
  coalescent event can occur only between lineages $x_1$ and
  $x_1$. The left panel illustrates this case. Prior to time $t_2$,
  coalescent events can occur between any pair of lineages. The right
  panel illustrates the case in which $x_2$ and $z$ coalesce first.
  Bullets delineate the branches along which mutation would generate
  site pattern $x_1x_2$ (left panel) or $x_2z$ (right panel).}
\label{fig.wibfr}
\end{figure*}

Within a tree of populations, many different gene trees are
possible. For example, Fig.~\ref{fig.wibfr} illustrates two ways in
which the gene tree of lineages $x_1$, $x_2$, and $z$ can coalesce
within the same population tree. On the left, $x_1$ and $x_2$ coalesce
during the interval $(t_0,t_1)$ within population $X$. On the right,
no coalescent event occurs during this interval. Instead, the first
coalescent event occurs prior to time $t_2$, within the ancestral
population, $XYZ$.

When the coalescent event occurs within $(t_0, t_1)$, only one site
pattern can be produced: $x_1x_2$. But if it occurs earlier, before
time $t_2$, all three site patterns are equally likely. For this
reason, it is necessary to distinguish these cases when calculating
the probabilities of site patterns. Our analysis uses the method of
\citet{Durand:S-328-S162}, \citet{Durand:N-468-S169}, and
\citet{Durand:MBE-28-2239}. This section explains the general
principles.

\subsection{Expectation of a truncated exponential random variable} 
\label{sec.truncExp}

Let $t$ represent an exponential random variable with mean $K$.  We
are interested in the conditional expectation of $t$, given that $t$
is less than an arbitrary value, $z$. The mean can be written as
\[
K = F^{(0,z)} E[t | t<z] + S^{(0,z)} E[t | t>z]
\]
Where $S^{(0,z)} = e^{-z/K}$ is the probability that $t>z$, and
$F^{(0,z)} = 1-S^{(0,z)}$ is the probability that $t<z$. The
memoryless property of the exponential distribution implies that $E[t
  | t>z] = z + K$. Substituting and rearranging,
\begin{equation}
E[t | t<z] = K - z S^{(0,z)}/F^{(0,z)},
\label{eq.truncExp}
\end{equation}

\subsection{Coalescent waiting time under piecewise constant hazard}
\label{sec.T}

Suppose that population history is a sequence of epochs, numbered
backwards from the present. The $i$th epoch spans the interval $(t_i,
t_{i+1})$, where time is measured backwards from the present. Within
the $i$th epoch, the diploid population size (relative to $N_0$) is a
constant, $K_i$. For a single pair of lineages within interval $i$,
the hazard of a coalescent event is $1/K_i$ per coalescent time unit.

Consider a pair of lineages at time $x$, which lies within epoch
$i$. In other words, $t_i < x < t_{i+1}$. We are interested in the
expected time, $T_i^{(x)}$, until they coalesce. Either they coalesce within
epoch~$i$ (with probability $F_i^{(x,t_{i+1})}$), or they survive
into epoch~$i+1$ (with probability $S_i^{(x,t_{i+1})}$). In the
former case, the expected coalescence time is given by
Eqn.~\ref{eq.truncExp}. In the latter, it is
$t_{i+1}-x+T_{i+1}^{(t_{i+1})}$. This leads to
\begin{equation}
T_i^{(x)} = K_i + S_i^{(x,t_{i+1})}(T_{i+1}^{(t_{i+1})} - K_i)
\label{eq.G}
\end{equation}
We assume the final epoch is infinite. This implies that
$T_j^{(x)} = K_j$, if $j$ is the final epoch.  For other epochs,
Eqn.~\ref{eq.G} can be applied recursively to calculate
$T_i^{(x)}$. If population size is constant, $K_i = T_i^{(x)} = 1$
for all $i$.

If epoch $j$ represents the ancestral human population, then $K_j=1$,
as explained in section~\ref{sec.popsize}. We will often assume, in
addition, that $T_j^{(x)}=1$ in this population. This amounts to
assuming that no changes in population size occurred prior to the
ancestral human population. In the analyses of $p_D$ and $p_N$,
however, we extend the model farther back in time to include the
population ancestral to chimpanzees and humans.

\subsection{Coalescent event within an interval}
\label{sec.wi}

Consider the left panel of Fig.~\ref{fig.wibfr}, in which lineages
$x_1$ and $x_2$ coalesce during $(t_0, t_1)$, within population $X$.
This occurs with probability $P = F_X^{(t_0, t_1)}$, as explained in
section~\ref{sec.popsize}. In this case, the gene tree will be of form
$((x_1,x_2),z)$, as required by site pattern $x_1x_2$. But even when
this genealogy does arise, we cannot be sure that site pattern
$x_1x_2$ will appear in the data. That happens only when a mutation
falls on the branch delineated by bullets in the left
panel of Fig.~\ref{fig.wibfr}.

The probability of such a mutation depends on branch lengths. Let $w$
represent the time, within interval $(t_0,t_1)$, at which the
coalescent event occurs. In other words, the event occurs at time
$t_0+w$, measuring time backwards from the present. The expectation of
$w$ is given by Eqn.~\ref{eq.truncExp}.

Continuing into the past, we can trace the single lineage ancestral to
$x_1$ and $x_2$. At time $t_2$, it becomes part of population
\textit{XYZ}. After an additional $v$ units of time, it coalesces with
lineage $z$.  Here, $v$ is the coalescence time within population
\textit{XYZ} and has mean $T_{XYZ}^{(t_2)}$, as explained in
\ref{sec.T}.

Pattern~$x_1x_2$ is generated when a mutation occurs on the branch
that separates two coalescent events: one that joins $x_1$ and $x_2$,
and another that joins their common ancestor to lineage~$z$. This
branch is delineated by bullets in the left panel of
Fig.~\ref{fig.wibfr}.  Its length, $t_2 + v -t_0 - w$, has expectation 
\[
B = (t_1-t_0)/F_X^{(t_0, t_1)} + t_2-t_1+ T_{XYZ}^{(t_2)} - K_X 
\]
The probability that a mutation falls on this branch, producing site
pattern $x_1x_2$, is approximately $UB$, where $U$ is the
mutation rate defined in section~\ref{sec.popsize}.

Combining all this, we can calculate the probability that a coalescent
event falls within interval $(t_0,t_1)$ and gives rise to a site with
pattern $x_1x_2$. The probability is the product of $P$ (the
probability of the appropriate genealogy) and $UB$ (the probability of
a mutation on the resulting branch). Altogether, this equals $UPB$,
where
\[
PB = t_1-t_0 + (t_2-t_1+T_{XYZ}^{(t_2)} - K_X) F_X^{(t_0, t_1)}
\]
In a population of constant size, $K_X = T_{XYZ}^{(t_2)} = 1$.

\subsection{Coalescent event within the ancestral population}
\label{sec.bfr}

Consider now the case illustrated in the right panel of
Fig.~\ref{fig.wibfr}. No coalescent event occurs during
$(t_0,t_2)$---an event of probability
$S_X^{(t_0,t_1)}S_{XY}^{(t_1,t_2)}$. Consequently, all three lineages
enter the ancestral population, \textit{XYZ}. These can coalesce in any
order, so we are equally likely to see genealogies $((x_1,x_2),z)$,
$((x_1,z),x_2)$, and $(x_1,(x_2,z))$. The probability of any given
genealogy is $P = S_X^{(t_0,t_1)}S_{XY}^{(t_1,t_2)}/3$.

If the first pair of lineages coalesces at time $x > t_2$, the waiting
time until the second coalescent event is $B = T_{XYZ}^{(x)}$, as
explained in \ref{sec.T}. This case generates each of the three site
patterns with probability $UPB$, where
\[
P B = S_X^{(t_0,t_1)}S_{XY}^{(t_1,t_2)}T_{XYZ}^{(x)}/3
\]
A similar result was derived by \citet[Eqn.~6]{Hudson:G-131-509}.

If \textit{XYZ} is the ancestral human population, we will usually
assume that $T_{XYZ}^{(x)} = 1$, as discussed in \ref{sec.T}.

\section{Expected value of $\hat f$}
\label{sec.f}

\begin{table*}
  \caption{Contributions to each site pattern for estimators
    $\hat f$ and $Q$. Each row represents a different case---a
    different way in which coalescent events can be distributed within
    the population tree. $P_i$ is the probability of
    each such case, and $B_i$ is the conditionally expected length of
    the branch along which mutation would give rise to each site
    pattern. $P_i B_i$ is the contribution
    of the current case to the unconditionally expected branch
    length. ``Source'' is the population from which sample $y$
    derives. ``Coal.'' indicates the time interval containing the most
    recent coalescent event. ``Ref'' refers to equation numbers in
    \ref{sec.f}.}
\label{tab.formulae}
{\centering
\begin{tabular}{clll}
\hline
Source&Coal.&$P_i B_i$&Ref\\
\hline\hline
\multicolumn{3}{l}{\it Site Pattern~$ny$ only\/}\\
$N$
   & $(\delta, \lambda)$
   & $m_N(1-m_D)\bigl\{\lambda-\delta 
  + (1 - K_N)F_{N}^{(\delta,\kappa)}
  + (1 - K_{ND}) S_N^{(\delta,\kappa)} F_{ND}^{(\kappa,\lambda)}\bigr\}$
   & \ref{eq.new-ay-immigrant}\\
$D$
   & $(\kappa, \lambda)$
   & $m_D\bigl\{\lambda-\kappa + (1 - K_{ND})F_{ND}^{(\kappa,\lambda)}\bigr\}$
   & \ref{eq.new-y-D-ND}\\[1.5ex]
\multicolumn{3}{l}{\it Site Pattern~$xy$ only\/}\\
$XY$
   &$(\zeta, \lambda)$
   & $(1-m_N)(1-m_D)\bigl\{\lambda-\zeta 
   + (1 - K_{XY})F_{XY}^{(\zeta,\lambda)}\bigr\}$
   & \ref{eq.new-xy-native-XY}\\[1.5ex]
\multicolumn{3}{l}{\it All Three Site Patterns\/}\\
$N$
   &$>\lambda$
   & $m_N(1-m_D) S_N^{(\delta,\kappa)} S_{ND}^{(\kappa,\lambda)}/3$
   & \ref{eq.new-all-admixed-XYA}\\
$D$
   &$>\lambda$
   & $m_D S_{ND}^{(\kappa,\lambda)}/3$
   & \ref{eq.new-all-D-XYND}\\
$XY$
   &$>\lambda$
   & $(1-m_N)(1-m_D)S_{XY}^{(\zeta,\lambda)}/3$
   & \ref{eq.new-all-native-XYA}\\
\hline
\end{tabular}\\}
\end{table*}

$\hat f$ is defined in Eqn.~\ref{eq.f} and assumes the population
history in Fig.~\ref{fig.xynd}.  The numerator of $\hat f$ involves
four genomes, $x$, $y$, $n$, and $c$, sampled from two populations,
$X$ and $Y$, of modern humans, one archaic population, $N$, and the
chimpanzee population, $C$.  The chimpanzee population is used only to
determine which allele is derived and is not shown in
Fig.~\ref{fig.xynd}.  The Eurasian population received gene flow from
Neanderthals at time $\delta$ and from Denisovans at a later time,
$\alpha$. The first of these episodes replaced a fraction $m_N$ of the
Eurasian gene pool; the second replaced a fraction $m_D$.  The
remaining fraction $(1-m_N)(1-m_D)$ is ``native''---it derives from
the population, $XY$, of ancestral modern humans.

\begin{flushenum}
\item 
\emph{$y$ arrived by gene flow from $D$}
\begin{flushenum}
\item 
\emph{$y$ and $n$ coalesce within \textit{ND}, during $(\kappa,
  \lambda)$}.  Sites in this category contribute to pattern~$ny$ only.
  The argument in \ref{sec.wi} gives $P = m_D F_{ND}^{(\kappa,\lambda)}$;
  $B = (\lambda-\kappa)/F_{ND}^{(\kappa,\lambda)} + 1 - K_{ND}$; and
\begin{equation}
PB = m_D\bigl\{\lambda-\kappa + (1 - K_{ND})F_{ND}^{(\kappa,\lambda)}\bigr\}
\label{eq.new-y-D-ND}
\end{equation}

\item \emph{$y$ and $n$ coalesce in the ancestral human population.} This
  case can give rise to any of the three genealogies. The argument in
  \ref{sec.bfr} gives $P = m_D S_{ND}^{(\kappa,\lambda)}/3$; $B = 1$; and
\begin{equation}
PB = m_D S_{ND}^{(\kappa,\lambda)}/3
\label{eq.new-all-D-XYND}
\end{equation}
\end{flushenum}

\item\emph{$y$ arrived by gene flow from $N$.} 
\begin{flushenum}
\item\emph{$y$ and $n$ coalesce within $N$, during $(\delta,
    \kappa)$}.  Sites in this category contribute to pattern~$ny$
  only.  The argument in \ref{sec.wi} gives $P =
  m_N(1-m_D)F_{N}^{(\delta,\kappa)}$; $B= \lambda-\kappa + 1-K_{N} +
  (\kappa-\delta)/F_N^{(\delta,\kappa)}$; and 
\begin{multline}
PB = m_N(1-m_D)\bigl\{\kappa-\delta\\
 + (\lambda-\kappa + 1-K_{N})F_{N}^{(\delta,\kappa)}\bigr\}
\label{eq.new-ay-immigrant-N}
\end{multline}

\item\emph{$y$ and $n$ coalesce within \textit{ND}, during
    $(\kappa, \lambda)$}.  Contributes to
  pattern~$ny$ only.  The argument in \ref{sec.wi} gives
  $P = m_N(1-m_D)S_N^{(\delta,\kappa)} F_{ND}^{(\kappa,\lambda)}$; $B= 1 - K_{ND} +
  (\lambda-\kappa)/F_{ND}^{(\kappa,\lambda)}$; and
\begin{multline}
PB = m_N(1-m_D)S_N^{(\delta,\kappa)} \bigl\{\lambda-\kappa\\
    + (1 - K_{ND})F_{ND}^{(\kappa,\lambda)}\bigr\}
\label{eq.new-ay-immigrant-ND}
\end{multline}
The total contribution from these last two cases is obtained by
summing Eqns.~\ref{eq.new-ay-immigrant-N}
and~\ref{eq.new-ay-immigrant-ND}: 
\begin{multline}
PB = m_N(1-m_D)\bigl\{\lambda-\delta 
  + (1 - K_N)F_{N}^{(\delta,\kappa)}\\
  + (1 - K_{ND}) S_N^{(\delta,\kappa)} F_{ND}^{(\kappa,\lambda)}\bigr\}
\label{eq.new-ay-immigrant}
\end{multline}

\item\emph{$y$ and $n$ coalesce in the ancestral human population.}
  Contributes to all three site patterns. The argument in
  \ref{sec.bfr} gives $P = m_N(1-m_D) S_N^{(\delta,\kappa)}
  S_{ND}^{(\kappa,\lambda)}/3$; $B = 1$; and
\begin{equation}
PB =m_N(1-m_D) S_N^{(\delta,\kappa)} S_{ND}^{(\kappa,\lambda)} /3
\label{eq.new-all-admixed-XYA}
\end{equation}
\end{flushenum}

\item\emph{$y$ is native.} 
\begin{flushenum}
\item\emph{$x$ and $y$ coalesce in $XY$, during $(\zeta,
  \lambda)$.}  
Contributes to site pattern $xy$ only. The argument in
\ref{sec.wi} gives $P = (1-m_N)(1-m_D)F_{XY}^{(\zeta,\lambda)}$;
$B = 1 - K_{XY} + (\lambda-\zeta)/F_{XY}^{(\zeta,\lambda)}$; and
\begin{multline}
PB =(1-m_N)(1-m_D)\bigl\{\lambda-\zeta\\
 + (1 - K_{XY})F_{XY}^{(\zeta,\lambda)}\bigr\}
\label{eq.new-xy-native-XY} 
\end{multline}

\item\emph{$x$ and $y$ coalesce in the ancestral human population
  prior to $\lambda$.} 
Contributes to all three site patterns. The argument in \ref{sec.bfr}
gives $P =(1-m_N)(1-m_D)S_{XY}^{(\zeta,\lambda)}/3$; $B = 1$;
\begin{equation}
PB = (1-m_N)(1-m_D)S_{XY}^{(\zeta,\lambda)}/3
\label{eq.new-all-native-XYA}
\end{equation}
\end{flushenum}
\end{flushenum}
These results are summarized in table~\ref{tab.formulae}.

In this table, the only rows that contribute to pattern $nx$ are those
that contribute to all three site patterns. The expected value of
$I_{nx}$ is the sum of these rows times the product of $U$ (the
mutation rate per unit of coalescent time) and $L$ (the number of
nucleotide sites sampled):
\begin{multline}
E[I_{nx}] = \frac{UL}{3}
  \Bigl[m_N(1-m_D) S_N^{(\delta,\kappa)} S_{ND}^{(\kappa,\lambda)} + m_D S_{ND}^{(\kappa,\lambda)}\\
 + (1-m_N)(1-m_D) S_{XY}^{(\zeta,\lambda)}\Bigr]\label{eq.Inx}
\end{multline}
These same rows also contribute to site pattern $ny$, along with
several additional rows that contribute only to $ny$. Thus,
\begin{align}
E[I_{ny}] &= E[I_{nx}] + UL \Bigl[
   m_N(1-m_D) \bigl\{\lambda-\delta\nonumber\\
 & \quad + (1-K_N)F_N^{(\delta,\kappa)} 
  + (1-K_{ND})S_N^{(\delta,\kappa)} F_{ND}^{(\kappa,\lambda)}\bigr\}\nonumber\\
 &  \quad + m_D\bigl\{\lambda-\kappa 
  + (1-K_{ND})F_{ND}^{(\kappa,\lambda)}\bigr\}\Bigr]
\label{eq.Iny}
\end{align}
Similarly,
\begin{multline}
E[I_{xy}] = E[I_{nx}] 
   + UL (1-m_N)(1-m_D)\Bigl[\lambda-\zeta\\
   + (1-K_{XY}) F_{XY}^{(\zeta,\lambda)}\Bigr]
\label{eq.Ixy}
\end{multline}
Eqns.~\ref{eq.Inx}--\ref{eq.Iny} extend results derived by
\citet[Eqns.~3--4]{Durand:MBE-28-2239}.  

The excesses of
$I_{ny}$ and $I_{xy}$ over $I_{nx}$ are
\begin{subequations}
\begin{align}
\lefteqn{E[I_{ny} - I_{nx}] \propto m_N(1-m_D) \bigl\{\lambda-\delta}\hspace{2em}&\notag\\
 & + (1-K_N)F_N^{(\delta,\kappa)} 
  + (1-K_{ND})S_N^{(\delta,\kappa)}F_{ND}^{(\kappa,\lambda)}\bigr\}\notag\\
 & + m_D\bigl\{\lambda-\kappa  + (1-K_{ND})F_{ND}^{(\kappa,\lambda)}\bigr\}\label{eq.Inydiff}\\
\lefteqn{E[I_{xy} - I_{nx}] \propto (1-m_N)(1-m_D)\bigl\{\lambda-\zeta}\hspace{2em}&\notag\\
 & + (1-K_{XY}) F_{XY}^{(\zeta,\lambda)}\bigr\}
\label{eq.Ixydiff}
\end{align}
\end{subequations}
omitting the constant multiplier, $UL$.

The analysis of the denominator is identical to that of the numerator,
except that $m_N=1$ (because samples $n'$ and $n$ are both from the
same Neanderthal population), and $\delta$ (the time of gene flow) is
replaced by $\varepsilon$ (the age of the older of the two archaic
fossils) \citep[p.~169]{Durand:S-328-S162}. The symbol $m'_D$ will
represent the rate of Denisovan gene flow into Neanderthals. With
these changes, Eqn.~\ref{eq.Inydiff} becomes
\begin{align}
\lefteqn{E[J_{n'n} - J_{nx}] \propto (1-m'_D) \bigl\{\lambda-\varepsilon}\hspace{2em}&\notag\\
 & + (1-K_N)F_N^{(\delta,\kappa)} 
+ (1-K_{ND})S_N^{(\varepsilon,\kappa)}F_{ND}^{(\kappa,\lambda)}\bigr\}\notag\\
 & + m'_D\bigl\{\lambda-\kappa  + (1-K_{ND})F_{ND}^{(\kappa,\lambda)}\bigr\}\label{eq.Jnndiff}
\end{align}
The ratio of (\ref{eq.Inydiff}) to (\ref{eq.Jnndiff}) approximates
$E[\hat f]$ and is shown in Eqn.~\ref{eq.Ef1}.

\section{Expected value of $R_N$}
\label{sec.RN}

This method estimates the fraction, $m_N$, of Neanderthal gene flow into
Eurasians. It uses genomes sampled from five populations: African,
$A$, Eurasian, $E$, Neanderthal, $N$, Denisovan, $D$, and chimpanzee,
$C$. As usual, we use the corresponding lower-case letters to represent
genomes sampled from these populations. The method assumes that these
populations are related as shown in Fig.~\ref{fig.RNtree}.  Although
five populations are involved, only four are compared in the
numerator, and a different four are compared in the denominator. In
each panel of the figure, dashed lines indicate the sample that is
ignored.

$R_N$ is defined in Eqn.~\ref{eq.RN}, which is equivalent to the
definition of \citet[Eqn.~S8.3]{Patterson:N-468-S47}.  As discussed in
section~\ref{sec.results-RN}, we will consider a model in which the
Eurasian population received gene flow from Neanderthals at time
$\delta$ and from Denisovans at time $\alpha$. A fraction $m_N$ of the
Eurasian gene pool derives from Neanderthal admixture, and a fraction
$m_D$ derives from Denisovan admixture. The remaining fraction,
$(1-m_N)(1-m_D)$, is ``native''---it decends from the ancestral
population $AE$, of modern humans.

\subsection{Numerator of $R_N$}
\label{sec.RNnum}

The numerator of $R_N$ does not involve an archaic
genome. Nonetheless, we cannot ignore the Neanderthal and Denisovan
populations, because a fraction of Eurasian ancestry flows through
them. In the outline below, each item deals with one of the ways in
which coalescent events can be distributed within the population tree
shown in Fig.~\ref{fig.RNtree}. Results are summarized in
table~\ref{tab.RNnum}.

\begin{table}
\caption{Contributions to each site pattern for the numerator of
  $R_N$. ``Ref'' refers to equations in \ref{sec.RNnum}.}
\label{tab.RNnum}
{\centering
\begin{tabular}{lll}
\hline
Coal.& $P_i B_i$ & Ref\\
\hline\hline
\multicolumn{3}{l}{\it Site Pattern~$ae$ only:\/}\\
$(\zeta,\lambda)$
  & $(1-m_N)(1-m_D)\bigl\{\lambda-\zeta$\\
& \multicolumn{1}{r}{$\mbox{} + (1-K_{AE}) F_{AE}^{(\zeta,\lambda)}\bigr\}$}
  & \ref{eq.SNnum-ae-AE}\\[1ex]

\multicolumn{3}{l}{\it Site Pattern~$de$ only:\/}\\
$(\alpha,\kappa)$
  & $m_D\bigl\{\kappa-\alpha + (\lambda-\kappa+1-K_D)F_D^{(\alpha,\kappa)}\bigr\}$
  & \ref{eq.SNnum-de-D}\\
$(\kappa,\lambda)$
  & $m_D S_D^{(\alpha,\kappa)} \bigl\{\lambda-\kappa 
   + (1 - K_{ND})F_{ND}^{(\kappa,\lambda)}\bigr\}$
  & \ref{eq.SNnum-de-DND}\\
$(\kappa,\lambda)$
  & $m_N(1-m_D)\bigl\{\lambda-\kappa + (1-K_{ND})F_{ND}^{(\kappa,\lambda)}\bigr\}$
  & \ref{eq.SNnum-de-ND}\\[1ex]

\multicolumn{3}{l}{\it All Three Site Patterns:\/}\\
$>\lambda$
  & $m_D S_D^{(\alpha,\kappa)} S_{ND}^{(\kappa,\lambda)}/3$
  & \ref{eq.SNnum-all-eDenisovan}\\
$>\lambda$
  & $m_N(1-m_D) S_{ND}^{(\kappa,\lambda)}/3$
  & \ref{eq.SNnum-all-eNeanderthal}\\
$>\lambda$
  & $(1-m_N)(1-m_D) S_{AE}^{(\zeta,\lambda)} /3$
  & \ref{eq.SNnum-all-eNative}\\
\hline
\end{tabular}\\}
\end{table}

\noindent
\begin{flushenum} 
\item\emph{$e$ arrived by gene flow from Denisovans.}

\begin{flushenum} 
\item\emph{$e$ and $d$ coalesce during $(\alpha, \kappa)$, within
    $D$.}  Contributes to site pattern $de$ only. The argument in
  \ref{sec.wi} gives $P = m_D F_D^{(\alpha,\kappa)}$;
  $B=\lambda-\kappa + 1-K_D + (\kappa-\alpha)/F_D^{(\alpha,\kappa)}$.
\begin{equation}
PB = m_D\bigl\{\kappa-\alpha
      + (\lambda-\kappa + 1-K_D)F_D^{(\alpha,\kappa)}\bigr\}
\label{eq.SNnum-de-D}
\end{equation}

\item \emph{$e$ and $d$ coalesce during $(\kappa, \lambda)$, within
    \textit{ND}.}  Contributes to site pattern $de$ only. $P =
  m_D S_D^{(\alpha,\kappa)} F_{ND}^{(\kappa,\lambda)}$; $B=1-K_{ND} +
  (\lambda-\kappa)/F_{ND}^{(\kappa,\lambda)}$. Argument as in
  \ref{sec.wi}.
\begin{equation}
  PB = m_D S_D^{(\alpha,\kappa)} \bigl\{\lambda-\kappa
  + (1 - K_{ND})F_{ND}^{(\kappa,\lambda)}\bigr\}
\label{eq.SNnum-de-DND}
\end{equation}

\item \emph{First coalescent event is prior to time $\lambda$, in the
    ancestral human population.} Contributes to all three site
  patterns. $P = m_D S_D^{(\alpha,\kappa)}
  S_{ND}^{(\kappa,\lambda)}/3$; $B=1$. Argument as in \ref{sec.bfr}.
\begin{equation}
PB = m_D S_D^{(\alpha,\kappa)} S_{ND}^{(\kappa,\lambda)}/3
\label{eq.SNnum-all-eDenisovan}
\end{equation}
\end{flushenum} 

\item
\emph{$e$ arrived by gene flow from Neanderthals.}

\begin{flushenum} 
\item \emph{$e$ and $d$ coalesce during $(\kappa,\lambda)$ within
    \textit{ND}.} Contributes to site pattern $de$ only. $P =
  m_N (1-m_D) F_{ND}^{(\kappa,\lambda)}$; $B = 1-K_{ND} +
  (\lambda-\kappa)/F_{ND}^{(\kappa,\lambda)}$. 
\begin{equation}
PB = m_N(1-m_D)\bigl\{\lambda-\kappa
     + (1-K_{ND})F_{ND}^{(\kappa,\lambda)}\bigr\} 
\label{eq.SNnum-de-ND}
\end{equation}
Argument as in \ref{sec.wi}.
\item
\emph{First coalescent event is prior to $\lambda$, in the
  ancestral population.} Contributes to all three site patterns.
  $P = m_N(1-m_D) S_{ND}^{(\kappa,\lambda)}/3$; $B=1$. 
\begin{equation}
PB = m_N(1-m_D) S_{ND}^{(\kappa,\lambda)}/3
\label{eq.SNnum-all-eNeanderthal}
\end{equation}
Argument as in \ref{sec.bfr}.
\end{flushenum} 

\item
\emph{$e$ is native.}
\begin{flushenum} 
\item \emph{$a$ and $e$ coalesce during $(\zeta,\lambda)$ within
    $AE$.} Contributes to site pattern $ae$ only. $P =
  (1-m_N)(1-m_D)F_{AE}^{(\zeta,\lambda)}$; $B=1-K_{AE} +
  (\lambda-\zeta)/F_{AE}^{(\zeta,\lambda)}$.  
\begin{multline}
  PB = (1-m_N)(1-m_D)\bigl\{\lambda-\zeta\\
  + (1-K_{AE}) F_{AE}^{(\zeta,\lambda)}\bigr\}
\label{eq.SNnum-ae-AE}
\end{multline}
Argument as in \ref{sec.wi}.

\item
\emph{First coalescent event is prior to $\lambda$, in the
  ancestral population.} Contributes to all three site patterns.
  $P = (1-m_N)(1-m_D) S_{AE}^{(\zeta,\lambda)}/3$; $B=1$. 
\begin{equation}
PB = (1-m_N)(1-m_D) S_{AE}^{(\zeta,\lambda)}/3
\label{eq.SNnum-all-eNative}
\end{equation}
Argument as in \ref{sec.bfr}.
\end{flushenum} 
\end{flushenum} 

The last three rows of table~\ref{tab.RNnum} refer to cases in which the
first coalescent event occurs in the population ancestral to all
humans, including archaics. These contribute equally to all three site
patterns. Consequently, they disappear from the expected value of the
numerator of $R_N$:
\begin{align}
\lefteqn{E[I_{de} - I_{ad}]\propto}\hspace{1em}\notag\\ 
&m_N(1-m_D)\{\lambda-\kappa + (1-K_{ND})F_{ND}^{(\kappa,\lambda)}\}\notag\\
&+ m_D\{\lambda-\alpha +(1-K_D)F_D^{(\alpha,\kappa)} \notag\\
&\qquad\qquad + (1-K_{ND})S_D^{(\alpha,\kappa)}F_{ND}^{(\kappa,\lambda)}\}
\label{eq.ERNnum}
\end{align}
This expression omits the constant multiplier, $UL$.

\subsection{Denominator of $R_N$}
\label{sec.RNdenom}

The denominator of $R_N$ includes a Neanderthal genome but does not
include a Eurasian. For this reason, one-directional gene flow from
archaics to Eurasians does not affect the calculation. The analysis in
this case is very simple, so we will not record results in a table.
\noindent
\begin{flushenum} 
\item
\emph{$n$ and $d$ coalesce during $(\kappa,\lambda)$ within
  \textit{ND}.} Contributes to $dn$ only. $P = F_{ND}$;
$B = 1-K_{ND} + (\lambda-\kappa)/F_{ND}^{(\kappa,\lambda)}$; 
\begin{equation}
PB =\lambda-\kappa + (1-K_{ND})F_{ND}^{(\kappa,\lambda)}
\label{eq.SNdenom-dn-DN}
\end{equation}
Argument as in \ref{sec.wi}.

\item
\emph{First coalescent event is prior to $\lambda$ in the
  ancestral human population.} Contributes to all three site patterns.
  $P=S_{ND}^{(\kappa,\lambda)}/3$; $B = 1$; 
\begin{equation}
PB = S_{ND}^{(\kappa,\lambda)}/3
\label{eq.SNdenom-all-ancestral}
\end{equation}
Argument as in \ref{sec.bfr}.
\end{flushenum} 
\noindent
These results imply that the denominator of $R_N$ has expected value
\begin{equation}
E[J_{dn} - J_{ad}] \propto \lambda-\kappa + (1-K_{ND})F_{ND}^{(\kappa,\lambda)}
\label{eq.ERNdenom}
\end{equation}
Using a ratio of expectations to approximate the expectation of a
ratio, $E[R_N]$ is equal to the ratio of Eqns.~\ref{eq.ERNnum}
and~\ref{eq.ERNdenom}.

\section{Expected value of $R_D$}
\label{sec.RD}

\citet[Eqn.~S8.5]{Patterson:N-468-S47} also define a second statistic,
their $R_{\text{Denisova}}$, which we refer to as $R_D$. It assumes
the setup shown in Fig.~\ref{fig.RDtree} and can be defined as in
Eqn.~\ref{eq.RD}.  The numerator of $R_D$ compares genomes sampled
from populations $S$, $E$, $V$, and $C$. There are three site
patterns---$es$, $ev$, and $sv$---that conform to the requirement that
the chimpanzee carry the ancestral allele, and that the derived allele
be present in two other populations. The numerator of $R_D$ compares
two of these. The denominator of $R_D$ compares genomes from $S$, $Y$,
$D$, and $C$, so the legal three site patterns are $sy$, $ds$, and
$dy$. Of these, the denominator compares $ds$ and $sy$.

This model assumes that archaic DNA entered modern populations via two
episodes of gene flow. First, at time $\delta$, there was gene flow
from Neanderthals into the ancestor of all Eurasian populations. Just
after this event, the fraction of Neanderthal DNA in modern human
populations was $m_N$. Later, at time $\alpha$, the Melanesian
population experienced a second episode of gene flow, this time from
Denisovans. A fraction $m_D$ of Melanesian genes derive from
Denisovans via this episode of gene flow. Of the remaining fraction,
$1-m_D$, a fraction $m_N$ derives from Neanderthals, and another
fraction, $1-m_N$, is native. Thus, the DNA of modern Melanesians can
be divided into three components: a fraction $m_D$ is Denisovan, a
fraction $m_N(1-m_D)$ is Neanderthal, and a fraction $(1-m_N)(1-m_D)$
is native. In other Eurasian populations, there are only two
components: a fraction $m_N$ is Neanderthal, and the remaining
fraction $1-m_N$ is native.

\subsection{Numerator of $R_D$}
\label{sec.RDnum}

\begin{table*}
\caption{Contributions to each site pattern for the numerator of
  $R_D$. ``Ref'' refers to explanations in numbered paragraphs of
  \ref{sec.RDnum}.}
\label{tab.RDnum}
{\centering
\begin{tabular}{lll}
\hline
Coalescence& $P_i B_i$ & Ref\\
\hline\hline
\multicolumn{3}{l}{\it Site Pattern~$es$ only:\/}\\
$(\eta,\lambda)$
  &$m_D(1-m_N)\{\lambda-\eta + (1-K_{\text{\it SYEV}}) F_{\text{\it SYEV}}^{(\eta,\lambda)}$
  & \ref{eq.RDnum-es-eNative-mAdmixed-SEM}\\
$(\eta, \lambda)$
  &$m_N(1-m_N)(1-m_D)S_{EV}^{(\beta,\delta)}\bigl\{\lambda-\eta + (1 -
  K_{\text{\it SYEV}})F_{\text{\it SYEV}}^{(\eta,\lambda)}\bigr\}$ 
  &\ref{eq.RDnum-es-EMS}\\[1ex]

\multicolumn{3}{l}{\it Site Pattern~$ev$ only:\/}\\
$(\kappa,\lambda)$
  &$m_N m_D\{\lambda-\kappa + (1-K_{ND})F_{ND}^{(\kappa,\lambda)}$
  & \ref{eq.RDnumMMwi}\\
$(\beta,\delta)$
  &$(1-m_D)\bigl[\delta -\beta
  + F_{EV}^{(\beta,\delta)}\bigl\{\eta-\delta 
  + m_N(\lambda-\eta + 1-T_{\text{\it SYEV}}^{(\eta)}) 
  + T_{\text{\it SYEV}}^{(\eta)} - K_{EV}\bigr\}\bigr]$
&\ref{eq.RDnum-em-betagamma}\\
$(\delta, \kappa)$
  &$m_N^2(1-m_D)S_{EV}^{(\beta,\delta)} \bigl\{\kappa-\delta
 + (\lambda-\kappa + 1-K_N)F_N^{(\delta,\kappa)}\bigr\}$
  &\ref{eq.RDnum-em-N}\\
$(\kappa, \lambda)$
  &$m_N^2(1-m_D) S_{EV}^{(\beta,\delta)} S_N^{(\delta,\kappa)} \bigl\{
\lambda-\kappa + (1 - K_{ND})F_{ND}^{(\kappa,\lambda)}\bigr\}$
  &\ref{eq.RDnum-em-ND}\\
$(\delta,\eta)$
  &$(1-m_N)^2(1-m_D) S_{EV}^{(\beta,\delta)}\bigl\{\eta - \delta +
(T_{\text{\it SYEV}}^{(\eta)} - K_{EV})F_{EV}^{(\delta,\eta)}\bigr\}$  
  &\ref{eq.RDnum-em-eNative-mNative}\\[1ex]

\multicolumn{3}{l}{\it Site Pattern~$sv$ only:\/}\\
$(\eta,\lambda)$
  &$m_N(1-m_N)(1-m_D) S_{EV}^{(\beta,\delta)}\bigl\{\lambda-\eta + (1
  - K_{\text{\it SYEV}})F_{\text{\it SYEV}}^{(\eta,\lambda)}\bigr\}$ 
  &\ref{eq.ms-mNative-eNeanderthal-EMS}\\[1ex]

\multicolumn{3}{l}{\it All Three Site Patterns:\/}\\
$>\lambda$
  &$m_N m_D S_{ND}^{(\kappa,\lambda)}/3$
  &\ref{eq.RDnumMMbefore}\\
$>\lambda$
  &$m_D(1-m_N) S_{\text{\it SYEV}}^{(\eta,\lambda)}/3$
  &\ref{eq.RDnum-all-Enative-mDenisovan-ancestral}\\
$>\lambda$
  &$m_N^2(1-m_D) S_{EV}^{(\beta,\delta)} S_N^{(\delta,\kappa)}
S_{ND}^{(\kappa,\lambda)}/3$
  &\ref{eq.RDnum-all-eNeanderthal-mNeanderthal-ancestral}\\
$>\lambda$
  &$m_N(1-m_N)(1-m_D) S_{EV}^{(\beta,\delta)} S_{\text{\it SYEV}}^{(\eta,\lambda)}/3$
  &\ref{eq.RDnum-all-eNative-mNeanderthal}\\
$>\lambda$
  &$m_N(1-m_N)(1-m_D) S_{EV}^{(\beta,\delta)} S_{\text{\it SYEV}}^{(\eta,\lambda)}/3$
  &\ref{eq.RDnum-all-mNative-eNeanderthal-ancestral}\\
$>\eta$
  &$\frac{1}{3}(1-m_N)^2(1-m_D)S_{EV}^{(\beta,\eta)}\bigl[K_{\text{\it SYEV}}
+ \frac{3}{2}(1-K_{\text{\it SYEV}})S_{\text{\it SYEV}}^{(\eta,\lambda)}
 - \frac{1}{2}(1-K_{\text{\it SYEV}}) e^{-3(\lambda-\eta)/K_{\text{\it SYEV}}}\bigr]$
  &\ref{eq.RDnum-all-mNative-eNative-SYEVorAncestral}\\
\hline
\end{tabular}\\}
\end{table*}

Each item below deals with one way in which coalescent events can be
distributed in the population tree shown in the left panel of
Fig.~\ref{fig.RDtree}. Results are collected in table~\ref{tab.RDnum}.
\noindent
\begin{flushenum} 
\item
\emph{$v$ derives from Denisovan gene flow.}

\begin{flushenum} 
\item
\emph{$e$ derives from Neanderthal gene flow.}

\begin{flushenum} 
\item \emph{$v$ and $e$ coalesce during $(\kappa, \lambda)$, within
    \textit{ND}}. This case contributes to site pattern $ev$ only.  $P=m_N
  m_D F_{ND}^{(\kappa,\lambda)}$;
  $B=(\lambda-\kappa)/F_{ND}^{(\kappa,\lambda)} + 1-K_{ND}$.
\begin{equation}
PB = m_N m_D \bigl\{\lambda-\kappa
    + (1-K_{ND}) F_{ND}^{(\kappa,\lambda)} \bigr\}
\label{eq.RDnumMMwi}
\end{equation}
Argument as in \ref{sec.wi}.
\item
\emph{First coalescent event is prior to $\lambda$ in the ancestral
  population}. Contributes to all three site patterns.
$P=m_N m_D S_{ND}^{(\kappa,\lambda)}/3$; $B=1$.
\begin{equation}
PB = m_N m_D S_{ND}^{(\kappa,\lambda)}/3
\label{eq.RDnumMMbefore}
\end{equation}
Argument as in \ref{sec.bfr}.
\end{flushenum} 

\item
\emph{$e$ is native.}

\begin{flushenum} 
\item
\emph{$e$ and $s$ coalesce during $(\eta, \lambda)$, within
  \textit{SYEV}.} Contributes only to site pattern $es$.
$P = m_D(1-m_N) F_{\text{\it SYEV}}^{(\eta,\lambda)}$;
$B = (\lambda-\eta)/F_{\text{\it SYEV}}^{(\eta,\lambda)} + 1-K_{\text{\it SYEV}}$.
\begin{multline}
PB = m_D(1-m_N) \bigl\{\lambda-\eta\\
    + (1-K_{\text{\it SYEV}}) F_{\text{\it SYEV}}^{(\eta,\lambda)}\bigr\}
\label{eq.RDnum-es-eNative-mAdmixed-SEM}
\end{multline}
Argument as in \ref{sec.wi}.

\item
\emph{$e$ and $s$ coalesce prior to $\lambda$, within ancestral
  population.} Contributes to all three site patterns.
$P = m_D(1-m_N) S_{\text{\it SYEV}}^{(\eta,\lambda)}/3$;
$B=1$.
\begin{equation}
PB = m_D(1-m_N) S_{\text{\it SYEV}}^{(\eta,\lambda)}/3
\label{eq.RDnum-all-Enative-mDenisovan-ancestral}
\end{equation}
Argument as in \ref{sec.bfr}.
\end{flushenum} 
\end{flushenum} 

\item\emph{$v$ not from Denisova, and
    coalesces with $e$ during $(\beta,\delta)$ within $EV$.}
\label{it.RDnum-em-betagamma}
Contributes to $ev$ only.  The frequency of such sites is $P =
(1-m_D)F_{EV}^{(\beta,\delta)}$. The branch length is
\[ 
m_N (\lambda + v) + (1-m_N) (\eta + x) -\beta - w
\]
where $v$ is the coalescence time in the ancestral human population
and therefore has expectation~1. Variable $x$ is the coalescence time
in \textit{SYEV} and has expectation $T_{\text{\it SYEV}}^{(\eta)}$, as explained
in \ref{sec.T}. Variable $w$ is the conditional coalescence
time, given that coalescence occurs within $(\beta,\delta)$.  Its
expection is
\[
E[w | w < \delta-\beta] = K_{EV} -
(\delta-\beta)S_{EV}^{(\beta,\delta)} / F_{EV}^{(\beta,\delta)}
\]
as explained in \ref{sec.truncExp}.

Assembling these pieces,
\begin{multline*}
B = \eta - \delta + (\delta-\beta)/F_{\text{\it SYEV}}^{(\eta,\lambda)}\\
 + m_N\bigl\{\lambda-\eta + 1-T_{\text{\it SYEV}}^{(\eta)}\bigr\} 
 + T_{\text{\it SYEV}}^{(\eta)} - K_{EV}
\end{multline*}
and
\begin{align}
PB &= (1-m_D)\bigl[\delta -\beta
 + F_{EV}^{(\beta,\delta)}\bigl\{\eta-\delta + m_N(\lambda-\eta)\notag\\  
 &\quad + m_N(1-T_{\text{\it SYEV}}^{(\eta)}) + T_{\text{\it SYEV}}^{(\eta)} - K_{EV}\bigr\}\bigr]
\label{eq.RDnum-em-betagamma}
\end{align}

\item
\emph{$v$ derives from $N$.}

\begin{flushenum} 
\item
\emph{$e$ derives from $N$.}
\begin{flushenum} 
\item \emph{$e$ and $v$ coalesce during $(\delta, \kappa)$ within
    $N$.}  Contributes to $ev$ only.  $P = m_N^2 (1-m_D)
  S_{EV}^{(\beta,\delta)} F_N^{(\delta,\kappa)}$; $B = \lambda -
  \kappa + 1-K_N + (\kappa-\delta)/F_N^{(\delta,\kappa)}$. 
\begin{multline}
PB = m_N^2(1-m_D)S_{EV}^{(\beta,\delta)} \bigl\{\kappa-\delta\\
+ (\lambda-\kappa + 1-K_N)F_N^{(\delta,\kappa)}\bigr\}
\label{eq.RDnum-em-N}
\end{multline}
Argument as in \ref{sec.wi}.

\item 
\emph{$e$ and $v$ coalesce during $(\kappa,\lambda)$ within $DN$.}
Contributes to $ev$ only.
$P = m_N^2(1-m_D) S_{EV}^{(\beta,\delta)} S_N^{(\delta,\kappa)} F_{ND}^{(\kappa,\lambda)}$;
$B = 1 - K_{ND} + (\lambda-\kappa)/F_{ND}^{(\kappa,\lambda)}$.
\begin{multline}
PB = m_N^2(1-m_D) S_{EV}^{(\beta,\delta)} S_N^{(\delta,\kappa)}
\bigl\{\lambda-\kappa\\
 + (1 - K_{ND})F_{ND}^{(\kappa,\lambda)}\bigr\}
\label{eq.RDnum-em-ND}
\end{multline}
Argument as in \ref{sec.wi}.

\item 
\emph{First coalescent event is prior to $\lambda$ in the ancestral
  population.} Contributes to all three site patterns.
$P = m_N^2(1-m_D) S_{EV}^{(\beta,\delta)} S_N^{(\delta,\kappa)}
S_{ND}^{(\kappa,\lambda)}/3$; $B = 1$.
\begin{equation}
PB = m_N^2(1-m_D) S_{EV}^{(\beta,\delta)} S_N^{(\delta,\kappa)}
S_{ND}^{(\kappa,\lambda)}/3
\label{eq.RDnum-all-eNeanderthal-mNeanderthal-ancestral}
\end{equation}
Argument as in \ref{sec.bfr}.
\end{flushenum} 

\item \emph{$e$ is native.}
\begin{flushenum} 
\item
\emph{$e$ and $s$ coalesce during $(\eta, \lambda)$ within
  \textit{EVS}.} Contributes to $es$ only.  $P =
m_N(1-m_N)(1-m_D)S_{EV}^{(\beta,\delta)}F_{\text{\it SYEV}}^{(\eta,\lambda)}$; $B =
1 - K_{\text{\it SYEV}} + (\lambda-\eta)/F_{\text{\it SYEV}}^{(\eta,\lambda)}$. 
\begin{multline}
PB = m_N(1-m_N)(1-m_D)S_{EV}^{(\beta,\delta)}\bigl\{\\
\lambda-\eta + (1 - K_{\text{\it SYEV}})F_{\text{\it SYEV}}^{(\eta,\lambda)}\bigr\}
\label{eq.RDnum-es-EMS}
\end{multline}
Argument as in \ref{sec.wi}.

\item
\emph{First coalescent event is prior to $\lambda$ within ancestral
  population.}
Contributes to all three site patterns.
$P = m_N(1-m_N)(1-m_D)S_{EV}^{(\beta,\delta)}S_{\text{\it SYEV}}^{(\eta,\lambda)}/3$;
$B=1$.
\begin{equation}
PB = m_N(1-m_N)(1-m_D)
 S_{EV}^{(\beta,\delta)} S_{\text{\it SYEV}}^{(\eta,\lambda)}/3
\label{eq.RDnum-all-eNative-mNeanderthal}
\end{equation}
Argument as in \ref{sec.bfr}.
\end{flushenum} 
\end{flushenum} 

\item
\emph{$v$ is native.}

\begin{flushenum} 
\item
\emph{$e$ is from Neanderthal.}

\begin{flushenum} 
\item
\emph{$v$ and $s$ coalesce during $(\eta,\lambda)$ within
  \textit{EVS}.}
Contributes to $sv$ only.
$P = m_N(1-m_N)(1-m_D) S_{EV}^{(\beta,\delta)} F_{\text{\it SYEV}}^{(\eta,\lambda)}$;
$B = 1 - K_{\text{\it SYEV}} + (\lambda-\eta)/F_{\text{\it SYEV}}^{(\eta,\lambda)}$.
\begin{multline}
PB = m_N(1-m_N)(1-m_D) S_{EV}^{(\beta,\delta)}\bigl\{\\ 
\lambda-\eta + (1 - K_{\text{\it SYEV}})F_{\text{\it SYEV}}^{(\eta,\lambda)}\bigr\}
\label{eq.ms-mNative-eNeanderthal-EMS}
\end{multline}
Argument as in \ref{sec.wi}.

\item
\emph{First coalescent event is prior to $\lambda$ within ancestral
  population.}
Contributes to all three site patterns.
$P = m_N(1-m_N)(1-m_D) S_{EV}^{(\beta,\delta)} S_{\text{\it SYEV}}^{(\eta,\lambda)}/3$;
$B=1$.
\begin{equation}
PB = m_N(1-m_N)(1-m_D)
 S_{EV}^{(\beta,\delta)} S_{\text{\it SYEV}}^{(\eta,\lambda)}/3
\label{eq.RDnum-all-mNative-eNeanderthal-ancestral}
\end{equation}
Argument as in \ref{sec.bfr}.
\end{flushenum} 

\item
\emph{$e$ is native.}

\begin{flushenum} 
\item \emph{$e$ and $v$ coalesce during $(\delta,\eta)$ within $EV$.}
  Contributes to $ev$ only.  $P = (1-m_N)^2(1-m_D)
  S_{EV}^{(\beta,\delta)} F_{EV}^{(\delta,\eta)}$; $B =
  (\eta-\delta)/F_{EV}^{(\delta,\eta)} + T_{\text{\it SYEV}}^{(\eta)} -
  K_{EV}$.
\begin{multline}
PB = (1-m_N)^2(1-m_D) S_{EV}^{(\beta,\delta)}\bigl\{\\ 
\eta - \delta + (T_{\text{\it SYEV}}^{(\eta)} - K_{EV})F_{EV}^{(\delta,\eta)}\bigr\}
\label{eq.RDnum-em-eNative-mNative}.
\end{multline}
Argument as in \ref{sec.wi}.

\item 
\emph{First coalescent event is during $(\eta,\lambda)$ within \textit{SYEV}
  or during $(\lambda,\infty)$ within the ancestral population.}
Contributes to all three site patterns.
$P = (1-m_N)^2(1-m_D)S_{EV}^{(\beta,\eta)}/3$.
This case is unusual, in that, when we enter \textit{SYEV} at time $\eta$,
there are 3 lineages, and the hazard of a coalescent event is
$3/K_{\text{\it SYEV}}$. 
\begin{multline*}
B  = K_{\text{\it SYEV}} + \frac{3}{2}(1-K_{\text{\it SYEV}})e^{-(\lambda-\eta)/K_{\text{\it SYEV}}}\\
   - \frac{1}{2}(1-K_{\text{\it SYEV}}) e^{-3(\lambda-\eta)/K_{\text{\it SYEV}}}
\end{multline*}
Under constant population size, this collapses to $B=1$.
\begin{multline}
PB = \frac{1}{3}(1-m_N)^2(1-m_D)S_{EV}^{(\beta,\eta)}\bigl[K_{\text{\it SYEV}}\\ 
+ \frac{3}{2}(1-K_{\text{\it SYEV}})S_{\text{\it SYEV}}^{(\eta,\lambda)}\\
 - \frac{1}{2}(1-K_{\text{\it SYEV}}) e^{-3(\lambda-\eta)/K_{\text{\it SYEV}}}\bigr]
\label{eq.RDnum-all-mNative-eNative-SYEVorAncestral}
\end{multline}
\end{flushenum} 
\end{flushenum} 
\end{flushenum} 
The expected value of $I_{es}$ is $UL$ times the sum of terms in
table~\ref{tab.RDnum} that belong to site pattern $es$. A similar
statement holds for $I_{sv}$. Terms that are common to both site
patterns disappear from the expected difference, which therefore
equals
\begin{multline}
  E[I_{es} - I_{sv}] \propto m_D (1-m_N) \{\lambda-\eta\\
  + (1-K_{\text{\it SYEV}})F_{\text{\it SYEV}}^{(\eta,\lambda)}\}
\label{eq.ERDnum}
\end{multline}
ignoring the proportional multiplier, $UL$.  This is the expected
value of the numerator of $R_D$.

\subsection{Denominator of $R_D$}
\label{sec.RDdenom}

\begin{table}
\caption{Contributions to each site pattern for the denominator of
  $R_D$. ``Ref'' refers to explanations in numbered paragraphs of
  \ref{sec.RDdenom}.}
\label{tab.RDdenom}
{\centering
\begin{tabular}{lll}
\hline
Coalescence& $P_i B_i$ & Ref\\
\hline\hline
\multicolumn{3}{l}{\it Site Pattern~$sy$ only:\/}\\
$(\eta, \lambda)$
  &$\lambda-\eta + (1 - K_{\text{\it SYEV}})F_{\text{\it SYEV}}^{(\eta,\lambda)}$
  &\ref{eq.RDdenom-sy-SYEV}\\[1ex]

\multicolumn{3}{l}{\it All Three Site Patterns:\/}\\
$>\lambda$
  &$S_{\text{\it SYEV}}^{(\eta,\lambda)}/3$
  &\ref{eq.RDdenom-all-ancestral}\\
\hline
\end{tabular}\\}
\end{table}

The denominator of $R_D$ is based on the right panel of
Fig.~\ref{fig.RDtree}.
\begin{flushenum} 
\item\emph{$s$ and $y$ coalesce during $(\eta, \lambda)$ within
    \textit{SYEV}.} Contributes to $sy$ only. $P = F_{\text{\it SYEV}}^{(\eta,\lambda)}$;
  $B = 1 - K_{\text{\it SYEV}} + (\lambda-\eta)/F_{\text{\it SYEV}}^{(\eta,\lambda)}$.
\begin{equation}
PB = \lambda-\eta + (1 - K_{\text{\it SYEV}})F_{\text{\it SYEV}}^{(\eta,\lambda)}
\label{eq.RDdenom-sy-SYEV}
\end{equation}
Argument as in \ref{sec.wi}.

\item
\emph{First coalescent event occurs prior to $\lambda$ in the
  ancestral population, \textit{SYEVND}.} Contributes to all three site
  patterns. $P = S_{\text{\it SYEV}}^{(\eta,\lambda)}/3$; $B=1$.
\begin{equation}
PB = S_{\text{\it SYEV}}^{(\eta,\lambda)}/3
\label{eq.RDdenom-all-ancestral}
\end{equation}
  Argument as in \ref{sec.bfr}.
\end{flushenum} 
\noindent
These results are summarized in table~\ref{tab.RDdenom}. The expected
value of the denominator of $R_D$ is
\begin{equation}
E[J_{sy} - J_{ds}] \propto \lambda-\eta + (1 - K_{\text{\it SYEV}})F_{\text{\it SYEV}}^{(\eta,\lambda)}
\label{eq.ERDdenom}
\end{equation}
where we have ignored the proportional multiplier, $UL$.  The
expectation of $R_D$ is approximately the ratio of (\ref{eq.ERDnum})
to (\ref{eq.ERDdenom}), as shown in Eqn.~\ref{eq.ERD} of
section~\ref{sec.results-RD}.

\section{Expected value of $p_D$}
\label{sec.pD}

In deriving the expected value of $p_D$, \citet{Reich:AJH-89-516}
assume that changes in allele frequencies reflect genetic drift.  They
do not assume that drift has clock-like behavior. We will need
clock-like behavior, however, in order to deal with admixture from
Neanderthals and with non-simultaneous Denisovan admixture events.
This section derives the expected value of $p_D$ using a mutational
clock. 

Consider the population tree shown in figure~\ref{fig.pD}. In that
figure, $W$, $X$, and $Y$ represent modern human populations, $N$ and
$D$ represent archaic populations, and $C$ is chimpanzee. Population
$W$ is an outgroup within modern humans, and $X$ and $Y$ represent two
other modern human populations. The method assumes that $Y$ received
archaic gene flow, but $W$ and $X$ did not.  At locus $i$, let $w_i$,
$x_i$, $y_i$, and $d_i$ represent the frequency of the derived allele
in populations $W$, $X$, $Y$, and $D$.

\citet[Eqn.~1]{Reich:AJH-89-516} define $p_D$ as in our
Eqn.~\ref{eq.pDdef}. In the denominator of that expression, $y'_i$ is
the frequency within a fifth population, $Y'$. The method assumes that
$Y'$ occupies a position within the population tree that is similar to
that of $Y$. In other words, the tree of $W$, $X$, $Y'$, and $D$ has a
topology identical to that shown in figure~\ref{fig.pD}.  To model
ghost admixture, we allow for gene flow at time $\delta$ from the
Neanderthal population into population $Y$ and at time $\delta'$ into
population $Y'$.

Dropping subscripts, each term in the numerator of $p_D$ is $wx - wy -
xd + yd$.  Conditional on allele frequencies, $wx$ is the probability
that two random nucleotides, one drawn from $W$ and the other from
$X$, are both copies of the derived allele. The unconditional version
of this probability, $E[wx]$, is the probability that a mutation
occurred in a common ancestor of these two nucleotides. Because we
exclude sites at which the chimpanzee carries the derived allele, this
common ancestor must have lived after the separation with chimpanzees.

The probability of such a mutation depends on the length of the branch
separating two coalescent events: one between the two hominin lineages
and the other between the hominin and chimpanzee lineages.  If $\mu$
is the separation time of the hominin and chimpanzee populations, then
the mean coalescence time of the hominin and chimpanzee gene lineages
is $\mu+K_{CH}$, where the $K_{CH}$ represents the coalescence time
within the ancestral chimpanzee-hominin population. 

The coalescence time of the two hominin lineages is $\zeta+z$, where
$E[z] = T_{WXY}^{(\zeta)}$, as explained in \ref{sec.T}.  In
this notation,
\[
\begin{split}
E[wx] &\propto \mu + K_{CH} -\zeta - T_{WXY}^{(\zeta)}\\
E[wy] &\propto \mu + K_{CH}\\ 
 &\quad -(m_D + m_N(1-m_D))(\lambda + T_0^{(\lambda)})\\
 &\quad - (1-m_N)(1-m_D)(\zeta + T_{WXY}^{(\zeta)})\\ 
E[xd] &\propto \mu + K_{CH} - \lambda - T_0^{(\lambda)}\\
E[yd] &\propto \mu + K_{CH}\\
 &\quad - m_D(\alpha + T_D^{(\alpha)})\\
 &\quad - m_N(1-m_D)(\kappa + T_{ND}^{(\kappa)})\\
 &\quad -(1-m_N)(1-m_D)(\lambda + T_0^{(\lambda)})\\
\end{split}
\]
where $T_0^{(\lambda)}$ is the mean coalescence time for a pair of
lineages that enters the ancestral human population at time $\lambda$,
and we have omitted the constant multiplier $UL$. The numerator
of $p_D$ has expectation
\begin{align}
\lefteqn{E[wx - wy - xd + yd] \propto}\hspace{1em}\notag&\\
 & \qquad m_D(2\lambda - \zeta - \alpha) 
     + m_N(1-m_D)(2\lambda - \zeta - \kappa)\notag\\
 & \quad + m_D(2T_0^{(\lambda)} - T_{WXY}^{(\zeta)} - T_D^{(\alpha)})\notag\\
 & \quad + m_N(1-m_D)(2T_0^{(\lambda)} - T_{WXY}^{(\zeta)} - T_{ND}^{(\kappa)})
\label{eq.pDnum}
\end{align}
When population size is constant, all the $T_i$ equal unity, and the
last two lines of this expression disappear. That of the denominator
is similar, except that we have $\alpha'$, $m'_N$, and $m'_D$ in place
of $\alpha$, $m_N$, and $m_D$. The ratio of these approximates
$E[p_D]$ and is given in Eqn.~\ref{eq.EpD}.

\section{Expected value of $p_N$}
\label{sec.pN}

Dropping subscripts, each term in the numerator of Eqn.~\ref{eq.pNdef}
is $ad - dx - an + nx$, and each term in the denominator is $ad - dm -
an + mn$. The expected value of each product can be calculated as
explained in \ref{sec.pD}, using the assumptions in
Fig.~\ref{fig.pN}. Omitting the constant multiplier $UL$, the
expectations of the various products are 
\[
\begin{split}
E[ad] &\propto \mu + K_{CH} - \lambda - T_0^{(\lambda)}\\
E[dx] &\propto \mu + K_{CH} - m_D(\alpha+T_D^{(\alpha)})\\
      &\quad - m_N(1-m_D)(\kappa+T_{IMND}^{(\kappa)})\\
      &\quad - (1-m_N)(1-m_D)(\lambda+T_0^{(\lambda)})\\
E[an] &\propto \mu + K_{CH} - \lambda-T_0^{(\lambda)}\\
E[nx] &\propto \mu + K_{CH} - m_D(\kappa+T_{IMND}^{(\kappa)})\\
      &\quad - m_N(1-m_D)(\theta+T_{IMN}^{(\theta)})\\
      &\quad -(1-m_N)(1-m_D)(\lambda+T_0^{(\lambda)})\\
E[dm] &\propto \mu + K_{CH} - \kappa-T_{IMND}^{(\kappa)}\\
E[mn] &\propto \mu + K_{CH} - \theta-T_{IMN}^{(\theta)}
\end{split}
\]
The expected numerator of $p_N$ is
\begin{multline}
m_N(1-m_D)(\kappa-\theta + T_{IMND}^{(\kappa)} - T_{IMN}^{(\theta)})\\ 
- m_D(\kappa-\alpha+T_{IMND}^{(\kappa)}-T_D^{(\alpha)})
\end{multline}
The expected denominator is 
\[
\kappa-\theta + T_{IMND}^{(\kappa)} - T_{IMN}^{(\theta)}
\]
The ratio of these, given in Eqn.~\ref{eq.EpN}, approximates the
expectation of $p_N$.

\section*{Acknowledgements}

We are grateful for comments from 
Elizabeth Cashdan,
Henry Harpending,
Flora Jay,
and Montgomery Slatkin.

\section*{References}
\bibliographystyle{elsarticle-harv}
\bibliography{defs,arrpubs,molrec,tree,bone}

\end{document}


\title{Bias in Estimators of Archaic Admixture\\
Supplementary Materials}
\author{Alan R. Rogers\thanks{Dept.{} of Anthropology, 270~S 1400~E,
University of Utah, Salt Lake City, Utah
84112. rogers@anthro.utah.edu}
\and Ryan J. Bohlender}
\maketitle

\section{Validating theoretical formulas}
\label{sec.validation}

To check for errors in algebraic formulas, we developed software to
simulate the components of each estimator. These simulations generate
gene genealogies by running a coalescent process constrained by
assumptions involving the population tree (branch lengths as well as
topology), admixture events (timing as well as level of gene flow),
and the size of each population. 

Our goal was to estimate the expected values of the numerator and
denominator of each estimator---not the full sampling
distributions. Accordingly, we treated loci as independent and
simulated the genealogical process but not the mutational
process. This approach provides accurate estimates of expected values
but underestimates sampling variances.

For statistics, $\hat f$, $R_N$, $R_D$, $p_N$, and $Q$, we set
the fraction of primary admixture at 0.03. For $p_D$, we assumed that
$m_D$ equals 0.06 in the numerator but 0.03 in the denominator. All
other parameters were chosen at random. These randomized parameters
included population sizes, branch lengths in the population tree, the
timing of admixture events, and the level of ghost admixture.

\begin{figure*}
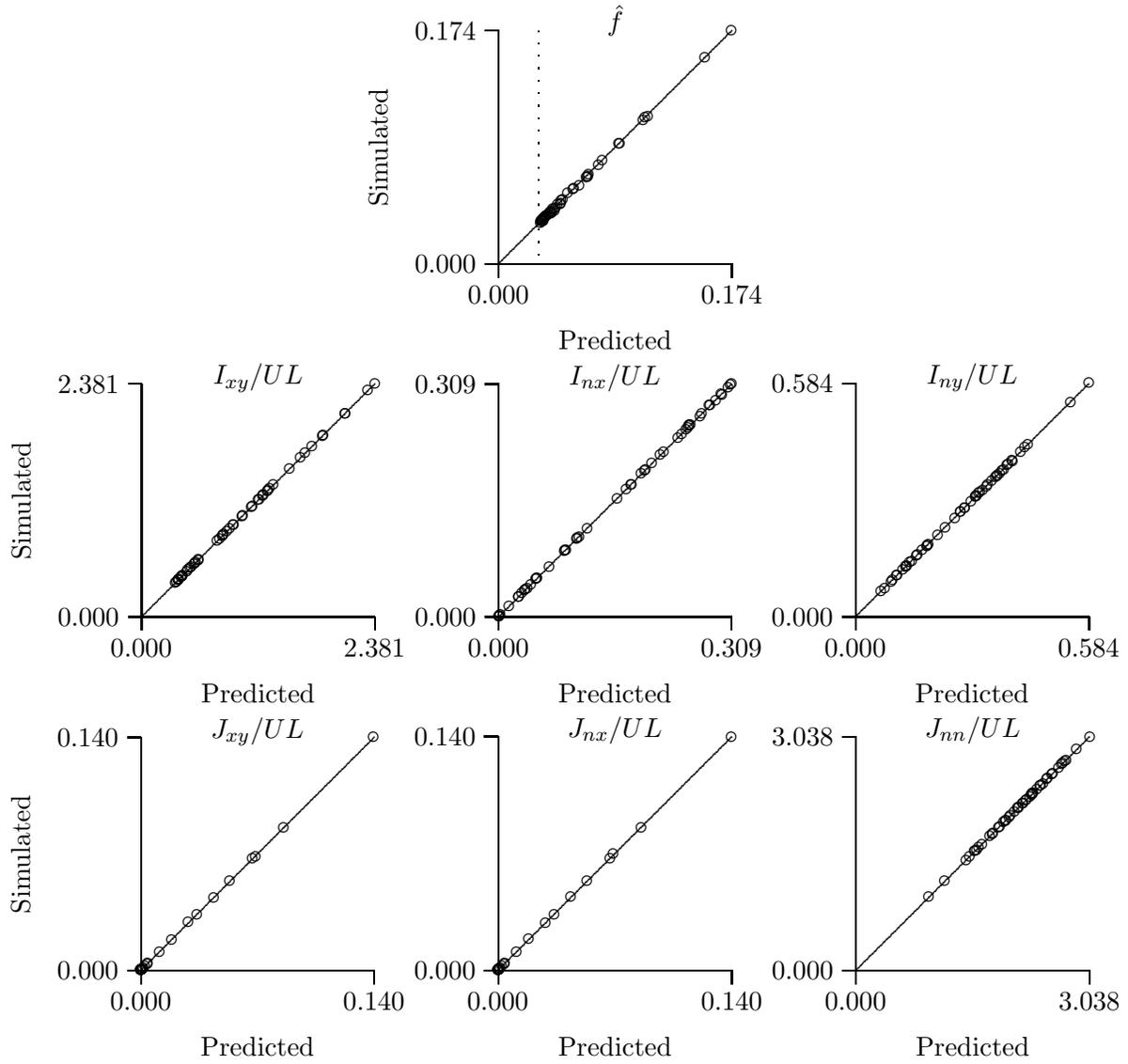

{\centering\input{figfhatsim}\\[1ex]
\input{figfhatIxy}\input{figfhatInx}\input{figfhatIny}\\[1ex]
\input{figfhatJxy}\input{figfhatJnx}\input{figfhatJnn}\\}
\caption{Simulated versus predicted values of $\hat f$ and its
  components. Simulations used $10^6$ iterations and assume
  $m_N=0.03$. All other parameters were chosen at random for each of
  50 simulations. Simulated and predicted values are equal along the
  solid 45-degree lines.}
\label{fig.fhatsim}
\end{figure*}

\begin{figure*}
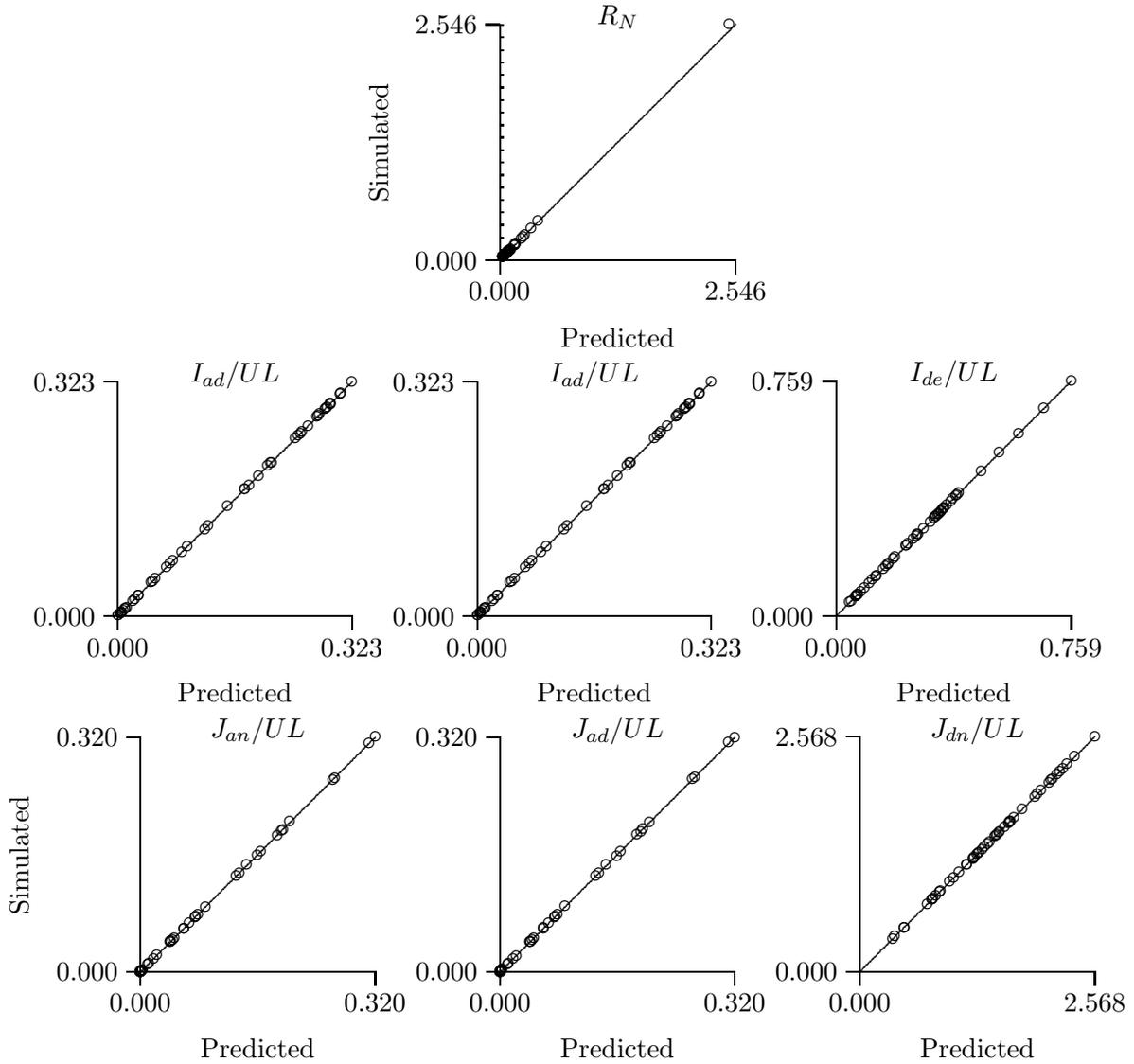

{\centering\input{figRNsim}\\[1ex]
\input{figRNIad}\input{figRNIad}\input{figRNIde}\\[1ex]
\input{figRNJan}\input{figRNJad}\input{figRNJdn}\\}
\caption{Simulated versus predicted values of $R_N$ and its
  components. Simulations used $10^6$ iterations and assume
  $m_N=0.03$. All other parameters were chosen at random for each of
  50 simulations. Simulated and predicted values are equal along the
  solid 45-degree lines.}
\label{fig.RNsim}
\end{figure*}

For each statistic, we considered 50 random sets of parameter
values. For each set, we estimated the statistic and its component
parts both by simulation and from the theoretical formulas given in
the main text. Simulated values are plotted against theoretical ones
in Figs.~\ref{fig.fhatsim}--\ref{fig.Qsim}. In each case, the points
fall close to a 45-degree line through the origin, indicating good
agreement between theory and simulation.

If these statistics were unbiased estimators of the level of primary
admixture, we would expect $\hat f$, $R_N$, $R_D$, $p_N$, and
$Q$ to cluster near the level, 0.03, of primary admixture. We would
expect $p_D$ to cluster near 2, the ratio of primary admixture in the
numerator to that in the denominator. (These targets are shown as
dotted vertical lines in the first panel of each figure below.)

This clustering is present, however, only in $R_D$ and $p_N$
(Figs.~\ref{fig.RDsim} and~\ref{fig.pNsim}). The values of the other
statistics are distributed along the 45-degree line, showing that they
are sensitive to variation in parameter values. There are several
outlying points in the graphs of the two $F_4$-ratio statistics, $p_D$
and $p_N$ (Figs.~\ref{fig.pDsim} and~\ref{fig.pNsim}). These
outliers are remarkable, because each point on these graphs is an average
across $10^8$ iterations. The sampling distributions of these
statistics must have very broad tails, at least for some sets of
parameter values.

The software used in these calculations is included in the
Supplementary Materials.

\begin{figure*}
{\centering\input{figRDsim}\\[1ex]
\input{figRDIes}\input{figRDIev}\input{figRDIsv}\\[1ex]
\input{figRDJsy}\input{figRDJds}\input{figRDJdy}\\}
\caption{Simulated versus predicted values of $R_D$ and its
  components. Simulations used $10^8$ iterations and assume
  $m_D=0.03$. All other parameters were chosen at random for each of
  50 simulations. Simulated and predicted values are equal along the
  solid 45-degree lines.} 
\label{fig.RDsim}
\end{figure*}

\begin{figure*}
{\centering\input{figpDsim}\input{figpDnum}\input{figpDdenom}\\[1ex]
\input{figpDwx}\input{figpDwy}\\[1ex]
\input{figpDxd}\input{figpDyd}\\}
\caption{Simulated versus predicted values of $p_D$ and its
  components. Simulations used $10^8$ iterations and assume $m_D=0.06$
  and $m'_D=0.03$. All other parameters were chosen at random for each
  of 50 simulations. Simulated and predicted values are equal along
  the solid 45-degree lines.}
\label{fig.pDsim}
\end{figure*}

\begin{figure*}
{\centering
\input{figpN}\\[1ex]
\input{figpNad}\input{figpNdx}\input{figpNan}\\[1ex]
\input{figpNnx}\input{figpNdm}\input{figpNmn}\\}
\caption{Simulated versus predicted values of $p_N$ and its
  components. Simulations used $10^8$ iterations and assume
  $m_N=0.03$. All other parameters were chosen at random for each of
  50 simulations. Simulated and predicted values are equal along 
  the solid 45-degree lines.}
\label{fig.pNsim}
\end{figure*}

\begin{figure*}
{\centering\input{figQsim}\\[1ex]
\input{figQIxy}\input{figQInx}\input{figQIny}\\}
\caption{Simulated versus predicted values of $Q$ and its
  components. Simulations used $10^6$ iterations and assume
  $m_N=0.03$. All other parameters were chosen at random for each of
  50 simulations. Simulated and predicted values are equal along the
  solid 45-degree lines.}
\label{fig.Qsim}
\end{figure*}